\documentclass[twocolumn,twocolappendix]{aastex63}

\usepackage{enumitem}
\usepackage{amsmath}
\usepackage[normalem]{ulem}  
\usepackage{multirow}
\usepackage{stackengine}
\usepackage{CJKutf8}
\usepackage{longtable}

\newcommand\redout{\bgroup\markoverwith
{\textcolor{red}{\rule[0.5ex]{2pt}{0.8pt}}}\ULon}
\newcommand\blueout{\bgroup\markoverwith
{\textcolor{blue}{\rule[0.5ex]{2pt}{0.8pt}}}\ULon}
\newcommand\schi{{\sc  H\,i}}%
\newcommand\schii{{\sc H\,ii}}%
\newcommand\rmht{\mathrm{H_2}}%
\newcommand\CO{{$\mathrm{^{12}CO}$}}%
\newcommand\COthtotw{{$\mathrm{^{12}CO}$ (3-2)}}%
\newcommand\COontoze{{$\mathrm{^{12}CO}$ (1-0)}}%
\newcommand\Rco{{$R_{31}$}}%
\newcommand\Rcotwo{{$R_{31,2}$}}%
\newcommand\um{{$\mu \mathrm{m}$}}%
\newcommand\kms{{\rm km\,s^{-1}}}%
\newcommand\Kkms{{\rm K\,km\,s^{-1}}}%
\newcommand\vlsr{{v_\mathrm{LSR}}}%
\newcommand\TAstar{{T_\mathrm{A}^*}}%
\newcommand\Tmb{{T_\mathrm{mb}}}%
\newcommand\wise{{\it WISE}}%
\newcommand\Msol{{\rm M_{\odot}}}
\newcommand\Lsol{{\rm L_{\odot}}}
\newcommand\Mgas{{M_\mathrm{gas}}}
\newcommand\Lbol{{L_\mathrm{bol}}}
\newcommand\dempseypaper{{Paper I}}%
\graphicspath{{./}{figures/}}

\begin{document}

\begin{CJK}{UTF8}{}
 \CJKfamily{mj}

\title{\COthtotw\ High-Resolution Survey (COHRS) of the Galactic Plane: Complete Data Release}


\author[0000-0001-8467-3736]{Geumsook Park (박금숙)}
\affil{Korea Astronomy and Space Science Institute, 776 Daedeokdae-ro, Yuseong-gu, Daejeon 34055, Republic of Korea}

\author[0000-0003-0141-0362]{Malcolm J. Currie}
\affil{East Asian Observatory, 660 N. A`oh\={o}k\={u} Place, HI 96720, USA}
\affil{RAL Space, STFC Rutherford Appleton Laboratory, Chilton, Didcot, Oxfordshire OX11 0QX, UK}

\author[0000-0003-1008-5477]{Holly S. Thomas}
\affil{Radio Telescope Data Center, Center for Astrophysics, Harvard \& Smithsonian, 60 Garden Street, Cambridge, MA 02138, USA}

\author[0000-0002-5204-2259]{Erik Rosolowsky}
\affil{Department of Physics, University of Alberta, 4-181 CCIS, Edmonton, AB T6G 2E1, Canada}

\author[0000-0002-5457-9025]{Jessica T. Dempsey}
\affil{East Asian Observatory, 660 N. A`oh\={o}k\={u} Place, HI 96720, USA}
\affil{ASTRON, Oude Hoogeveensedijk 4, 7991 PD Dwingeloo, The Netherlands}

\author[0000-0003-2412-7092]{Kee-Tae Kim (김기태)}
\affil{Korea Astronomy and Space Science Institute, 776 Daedeokdae-ro, Yuseong-gu, Daejeon 34055, Republic of Korea}

\author[0000-0002-3351-2200]{Andrew J. Rigby}
\affil{School of Physics and Astronomy, Cardiff University, Cardiff CF24 3AA, UK}
\affil{School of Physics and Astronomy, University of Leeds, Leeds LS2 9JT, UK}

\author{Yang Su}
\affil{Purple Mountain Observatory and Key Laboratory of Radio Astronomy, Chinese Academy of Sciences, Nanjing 210034, People's Republic of China}

\author[0000-0002-5881-3229]{David J. Eden}
\affil{Astrophysics Research Institute, Liverpool John Moores University, IC2, Liverpool Science Park, 146 Brownlow Hill, Liverpool L3 5RF, UK}
\affil{Armagh Observatory and Planetarium, College Hill, Armagh BT61 9DG, UK}

\author{Dario Colombo}
\affil{Max Planck Institute for Radioastronomy, Auf dem H\"{u}gel 69, D-53121 Bonn, Germany}
 
\author[0000-0002-6327-3423]{Harriet Parsons}
\affil{East Asian Observatory, 660 N. A`oh\={o}k\={u} Place, HI 96720, USA}

\author{Toby J. T. Moore}
\affil{Astrophysics Research Institute, Liverpool John Moores University, IC2, Liverpool Science Park, 146 Brownlow Hill, Liverpool L3 5RF, UK}

\begin{abstract}

We present the full data release of \COthtotw\ High-Resolution Survey (COHRS),
which has mapped the inner Galactic plane over the range of 
$9\fdg5 \le l \le 62\fdg3$ and $|b| \le 0\fdg5$.
The COHRS has been carried out using the Heterodyne Array Receiver Program (HARP)
on the 15\,m James Clerk Maxwell Telescope (JCMT) in Hawaii. 
The released data are smoothed to have a spatial resolution of 16\farcs6 and a velocity resolution of 0.635~$\kms$,
achieving a mean root-mean-square of $\sim 0.6$\,K on $\TAstar$. 
The COHRS data are useful for investigating
detailed three-dimensional structures of individual molecular clouds and 
large-scale structures such as spiral arms in the Galactic plane.
Furthermore, data from other available public surveys of different CO isotopologues and transitions
with similar angular resolutions to this survey,
such as FUGIN, SEDIGISM, and CHIMPS/CHIMPS2,
allow studying the physical properties of molecular clouds and comparing their states with each other.
In this paper, we report further observations on R2 and improved data reduction since the original COHRS release.
We discuss the characteristics of the COHRS data and present integrated-emission images and a position-velocity (PV) map of the region covered.
The PV map shows a good match with the spiral-arm traces from the existing CO and \schi\ surveys.
We also obtain and compare integrated one-dimensional distributions of \COontoze\ and (3-2) and those of star-forming populations to each other.

\end{abstract}


\section{Introduction} \label{sec:intro}
\end{CJK}

Carbon monoxide (CO) is essential for tracing the molecular interstellar medium (ISM) and understanding its physical properties.
Molecular hydrogen ($\rmht$) is the most abundant molecule in the interstellar medium,
but due to its symmetry and low molecular weight, 
there is a lack of transitions with energies that are excited under standard molecular ISM conditions with $T \sim 10$--100\,K.
As the second most abundant molecule,
CO is the best tracer of the bulk of the molecular ISM. 
Its rotational ($J$) transitions trace different conditions in the ISM.
Cold gas preferentially emits mostly in the $J=$ (1-0) line,
and warmer, denser gas radiates more in the higher-$J$ lines.
Different CO isotopologues reflect opacity effects.
$\mathrm{^{12}CO}$ is the most abundant CO isotopologue,
which produces strong emission lines from most of the molecular gas in the ISM.
$\mathrm{^{13}CO}$ and $\mathrm{C^{18}O}$ are less abundant and therefore fainter than $\mathrm{^{12}CO}$, but
trace optically thinner emission in most cases of low excitation temperature.
So $\mathrm{^{12}CO}$ traces where the molecular gas is,
while $\mathrm{^{13}CO}$ and $\mathrm{C^{18}O}$ provide a view of the optically thickest gas.
Thus, observations of different CO isotoplogues and transitions are complementary.

While continuum tracers blend emission along the line of sight, 
the high spectral resolution and Doppler effect of CO line emission allow us 
to identify discrete features in velocity that map to distance, although there is kinematic ambiguity within the Solar circle in position-position-velocity space.
Also, CO is often used as a mass tracer,
even though the emissivity per unit mass (that is, X-factor) is variable \citep{bolatto2013}.
Therefore, CO is useful for investigating detailed information about the morphological, physical, and kinematic properties of molecular clouds.

Molecular gas is predominantly distributed in the galactic disks of late-type galaxies, 
including the Milky Way. 
It traces large-scale galactic structures such as spiral arms, giant molecular clouds, and star-forming regions. 
Significant efforts have long been made to provide targeted surveys of the Galactic plane in several CO isotopologues and transitions by using a variety of facilities. 
Since the molecular gas in the Milky Way is centrally concentrated, 
CO surveys in the Milky Way usually focus on the first and fourth Galactic quadrants.
Table~\ref{tab:surveys} presents examples of Galactic CO surveys
in which the first quadrant is within the scope of the survey.
The early observations from 
the University of Massachusetts-Stony Brook (UMSB) survey \citep{sanders1986}, 
the CfA survey \citep{dame2001},
and the Boston University-Five College Radio Astronomy Observatory Galactic Ring Survey \citep{jackson2006}
provide \COontoze\ or $\mathrm{^{13}CO}$ (1-0) emission data.
In particular, since the CfA survey covers the entire Galactic plane,
it is still important for understanding the large-scale distribution of molecular gas,
although the angular resolution of the survey data is low by modern standards.

More-recent surveys simultaneously observe multiple lines of CO isotopologues
at moderate or much-higher angular resolutions.
For example, there are the FOREST Unbiased Galactic plane Imaging survey 
\citep[FUGIN;][]{umemoto2017};
the Milky Way Imaging Scroll Painting (MWISP) I and II projects
\citep[e.g., see][]{su2019, yuan2022};
and the structure, excitation, and dynamics of the inner Galactic\footnote{The Galactic region inside the Solar Circle is defined as the “inner” Galaxy.} interstellar medium (SEDIGISM) survey
\citep{schuller2017}.
In addition, two surveys that used the JCMT measure 
the (3-2) transition lines of three CO isotopologues ($\mathrm{^{12}CO/^{13}CO/C^{18}O}$).
For $\mathrm{^{12}CO}$, the CO high-Resolution Survey (COHRS),
of which the first release (R1) was presented by \citet[hereafter Paper I]{dempsey2013} and the second release (R2) is described in this paper, has been carried out.
For $\mathrm{^{13}CO}$ and $\mathrm{C^{18}O}$,
the CO Heterodyne Inner Milky Way Plane Survey \citep[CHIMPS;][]{rigby2016}
mapped part of regions covered by the COHRS.
The extension of this survey, CHIMPS2, is now proceeding
to extend into lower galactic longitudes, and to cover the Galactic center and an outer region \citep{eden2020}.
These recent surveys provide significantly improved CO data in angular resolution compared with the CfA survey.
Specifically, about 1000 times the area of the COHRS beam is equal to one CfA beam area.

\begin{deluxetable*}{lcllc ccccc}
\tabletypesize{\scriptsize}
\setlength{\tabcolsep}{0.04in}
\tablecaption{Examples of CO Surveys that Includes the First Galactic Quadrant \label{tab:surveys}}
\tablewidth{0pt}
\tablehead{
\colhead{} & \colhead{} & \multicolumn{2}{c}{} & \colhead{} &
\colhead{Angular} & \colhead{Angular} & \colhead{Velocity} & \colhead{RMS} & \colhead{} \vspace{-3.5mm}\\
\colhead{Survey Name} & \colhead{Molecule(s)} & \multicolumn{2}{c}{Survey Coverage} & \colhead{Telescope(s)} &
\colhead{Sampling} & \colhead{Resolution\tablenotemark{\footnotesize *}} & \colhead{Resolution\tablenotemark{\footnotesize *}} & \colhead{Sensitivity\tablenotemark{\footnotesize \#}} &\colhead{Ref.\tablenotemark{\footnotesize $\dagger$}} \vspace{-3mm} \\
\colhead{} & \colhead{} & \multicolumn{2}{c}{(\arcdeg)} & \colhead{} &
\colhead{(\arcsec)} & \colhead{(\arcsec)} & \colhead{($\kms$)} & \colhead{(K)} &  
}
\startdata
UMSB         & $\mathrm{^{12}CO~(J=1-0)}$                & $l=$ +8--+90               & $|b| \lesssim 1.05$   & FCRAO        & 180--360 & 45                & 0.65/1         & $\sim$0.6         & 1 \\
CfA          & $\mathrm{^{12}CO~(J=1-0)}$                & $l=$ $-$180--+180          & $|b| \lesssim$ 10--30 & CfA/Chile/NY & 225--450 & 504--528          & 0.65           & $\sim$0.1--0.4    & 2 \\ 
GRS          & $\mathrm{^{13}CO~(J=1-0)}$                & $l=$ +18--+55              & $|b| \lesssim$ 1      & FCRAO        & 22       & 46                & 0.21           & $\sim$0.1         & 3 \\
FUGIN        & $\mathrm{^{12}CO/^{13}CO/C^{18}O~(J=1-0)}$& $l=$ +10--+50, +198--+236  & $|b| \lesssim$ 1      & Nobeyama     & 8.5      & $\sim$14/$\sim$20 & 0.65/1.3       & $\sim$1.7/0.8/0.8 & 4 \\  
MWISP I+II\tablenotemark{\footnotesize $\ddagger$}                                                                                                                            
             & $\mathrm{^{12}CO/^{13}CO/C^{18}O~(J=1-0)}$& $l=$ +10--+240             & $|b| \lesssim$ 10.25  & PMO          & 15       & $\sim$50          & $\sim$0.16     & $\sim$0.2/0.1/0.1 & 5, 6 \\
SEDIGISM     & $\mathrm{^{13}CO/C^{18}O~(J=2-1)}$        & $l=$ $-$60--+18            & $|b| \lesssim$ 0.5    & APEX         & 15       & 28/30             & $\sim$0.1/0.25 & $\sim$0.4/0.4     & 7 \\
COHRS        & $\mathrm{^{12}CO~(J=3-2)}$                & $l=$ +9.5--+62.3           & $|b| \lesssim$ 0.5    & JCMT         & 6        & 13.8/16.6         & 0.42/0.635     & $\sim$0.8/0.8     & 8, 9 \\
CHIMPS       & $\mathrm{^{13}CO/C^{18}O~(J=3-2)}$        & $l=$ +28--+46              & $|b| \lesssim$ 0.5    & JCMT         & 7.3      & 15                & 0.055/0.5      & $\sim$0.6/0.7     & 10 \\
\multirow{2}{*}{CHIMPS2} & \multirow{2}{*}{$\mathrm{^{12}CO/^{13}CO/C^{18}O~(J=3-2)}$} & $l=$ $-5$--+28  & $|b| \lesssim$ 0.5    & \multirow{2}{*}{JCMT} & \multirow{2}{*}{\S} & \multirow{2}{*}{\S} & \multirow{2}{*}{\S} & \multirow{2}{*}{$\sim$1.0/.../...} & \multirow{2}{*}{11} \\
                         &                                                             & $l=$ +215--+225 & $-2$ $\leq b \leq$ 0  &                       &                     &                     &                     &  &  \\
\enddata
\tablenotetext{*}{If two values connected by a slash (/) are provided, the first is a raw value and the other is after smoothing.  Otherwise, a single entry is a raw value.}
\tablenotetext{\#}{Estimated RMS noise level ($\Tmb$) at a velocity channel width of 1~$\kms$ and a given original angular resolution. 
}
\tablenotetext{\dagger}{Reference(s):       
                  1- \citet{sanders1986},   
                  2- \citet{dame2001},      
                  3- \citet{jackson2006},   
                  4- \citet{umemoto2017},   
                  5- \citet{su2019},        
                  6- \citet{yuan2022},
                  7- \citet{schuller2017},  
                  8- \citet{dempsey2013},
                  9- this work,             
                  10- \citet{rigby2016},    
                  11- \citet{eden2020}.     
		         }
\tablenotetext{\ddagger}{The MWISP I project ($l=$ $+10\arcdeg$--$+230\arcdeg$, $|b| \lesssim$ 5.25\arcdeg) was completed over a 10-year period from 2011 to 2021.
The MWISP group plans to extend the Galactic latitude to $b=$ $\pm10.25\arcdeg$, i.e., the MWISP II project for $l=$ +10\arcdeg--+240\arcdeg\ and $|b| \lesssim$ 10.25\arcdeg\ region along the Galactic plane. 
}
\tablenotetext{\S}{For $\mathrm{^{12}CO}$ and $\mathrm{^{13}CO/C^{18}O}$ observations, CHIMPS observing strategies are followed that of COHRS, respectively.}
\end{deluxetable*}

COHRS observes \COthtotw, which is more optically thin
compared with the lower transition lines of the same isotopologue and 
is seen at a higher frequency, resulting in higher-resolution data given the same telescope diameter.
Compared with \COontoze, \COthtotw\ is excited in a warmer (energy above ground ($E_u$/$k_B$): 5.5\,K for $J=1$, 33\,K for $J=3$) and denser (critical density: $\sim 2 \times 10^3~{\rm cm^{-3}}$ for (1-0), $\sim 5 \times 10^4~{\rm cm^{-3}}$ for (3-2) in the optically thin regime) environment.
This transition traces molecular clouds, particularly gas that is likely to be more strongly associated with star-forming regions.
It is also an excellent tracer of outflow activity, 
which generally indicates the very early stages of star formation \citep{banerjee2006}.
Using the COHRS R1 data, \citet{li2018} established a catalog of high-mass outflows associated with ATLASGAL clumps,
resulting in the detection rate of 22\%. 
\citet{colombo2019} analysed integrated properties of molecular clouds
by applying the Spectral Clustering for Interstellar Molecular Emission Segmentation algorithm \citep{colombo2015}.
They identified 85,020 clouds and found that ~35\% of the classified clouds are located within spiral arms, assuming arm widths of 600~pc \citep{vallee2017}. 
They derived mass and size distributions showing the power-law relationship with spectral indices of $-$1.75 and $-$2.80, respectively, and the distributions are truncated at $\sim 3 \times 10^{6}~\Msol$ and $\sim70$~pc, respectively.

COHRS can be compared with other Galactic large-scale surveys at submillimeter and infrared wavelengths 
to study detailed structures of individual star-forming regions and large-scale structures in the Galactic plane. 
The existing continuum surveys covering the first Galactic quadrant are, for example, 
the APEX Telescope Large Area Survey of the Galaxy at 870~\um\ \citep[ATLASGAL;][]{schuller2009}, 
JCMT Galactic Plane Survey at 850~\um\ \citep[JPS;][]{moore2015, eden2017}, 
the Bolocam Galactic Plane Survey at 1.1~mm \citep[BGPS;][]{aguirre2011}, 
the Herschel infrared Galactic Plane Survey at 70--550~\um\ \citep[Hi-GAL;][]{molinari2010},
Wide-field Infrared Survey Explorer at 3.4--22~\um\ \citep[WISE;][]{wright2010}, 
Spitzer’s Galactic Legacy Infrared Mid-Plane Survey Extraordinaire at 3.6--8~\um\ \citep[GLIMPSE;][]{ benjamin2003, churchwell2009},
and MIPSGAL at 24 and 70~\um\ \citep{carey2009}.

This paper presents the {\it full} COHRS data covering 52.8~square degrees,
which is almost twice the first release,
by completing the overall planned latitudes in a more extensive longitude range than the previous one.
The R2 data are provided by mitigating off-position contamination mentioned in 
\dempseypaper\
and improved data reduction.
In Section~\ref{sec:obs}, we explain the COHRS observations and the general data-reduction procedure.
Section~\ref{sec:R2} provides the information of the second release of the full COHRS data
and how to access them online.
Section~\ref{sec:noise} describes the noise characteristics.
We present integrated position-position or position-velocity maps
and descriptions of one-dimensional distributions in Section~\ref{sec:results}.
Section~\ref{sec:example} shows examples of COHRS data for three star-forming regions.
In Section~\ref{sec:comp}, we analyze the one-dimensional distribution of COHRS data and compare it with those of other data, such as the lowest \CO\ transition and star-forming population.
We summarize main results in Section~\ref{sec:sum}.

\section{Observations and Data Reduction} \label{sec:obs}

\subsection{Observations}

COHRS is a spectral line survey mapping a strip of the Galactic plane in the first quadrant in \COthtotw\ line (345.786~GHz).
The survey covers a total of approximately 52.8 square degrees
in the region between $9\fdg5 \le l \le 62\fdg3$ and $|b| \le 0\fdg5$.
The observations were carried out with the Heterodyne Array Receiver Program \citep[HARP;][]{buckle2009}
on the 15\,m James Clerk Maxwell Telescope (JCMT) in Hawaii.

The target longitude range for the original survey was  $10\arcdeg \le l \le 65\arcdeg$. This was ultimately extended down to $l = 9\fdg5$ to follow the interesting structure seen in R1 around $l \sim 10\arcdeg$. The upper end of the range was truncated to $l \sim 62\arcdeg$; this decision was made in order to concentrate our remaining time maximizing the overlap between COHRS and the JPS, as well as re-observing the noisiest tiles from R1. We chose not to re-observe those tiles affected by contamination in the off-position. This decision was driven by time constraints and the desire to complete the survey area. We were confident that the effect of the contamination could be mitigated in post-processing.
We ultimately approached the issue of off-position signals by determining modal spectra and removing them during data reduction (see Section~\ref{sec:off-pos} for details).


The observations were taken over four semesters at JCMT during 2013--2014 and 2017--2018. 
As in previous observations, the observing time had been allocated as a mixture of PI time which includes some of CHIMPS2 time (proposal ID: M17BL004), Directors Discretionary Time, empty-queue or poor-weather backup time (Panel for the Allocation of Telescope Time (PATT) numbers: M13AU41, M13BN02, and M14AU09).
The data were collected in opacities ranging from $\tau_{225} \sim 0.06$ to $\tau_{225} \sim 0.31$.
The observational strategies follow those described in \dempseypaper.

HARP is a $4\times4$ array receiver with 16 superconductor-insulator-superconductor (SIS) heterodyne detectors\footnote{In fact, the number of operable detectors was 14. For more details, visit \url{https://www.eaobservatory.org/jcmt/instrumentation/heterodyne/harp/}.} arranged at intervals of 30\arcsec.
At the observing frequency, HARP has 
an angular resolution of 14\arcsec\ and a main-beam efficiency of $\eta_\mathrm{mb} = 0.61$.
The Auto-Correlation Spectral Imaging System \citep[ACSIS;][]{buckle2009} was used as a setting for an 1~GHz bandwidth and 2048 frequency channels, providing a frequency resolution of 0.488~MHz or a velocity resolution of 0.42~$\kms$.
The bandwidth was set to a velocity coverage of $-400~\kms < \vlsr < +400~\kms$.
The observations were taken in a position-switching raster (on-the-fly) mode with 1/2 array spacing. The bulk of the off-positions were measured above the Galactic plane with a latitude offset of $+2\fdg5$.

%
%


\subsection{Data Reduction} \label{sec:dr}

As with the original COHRS release, the observational data were reduced with
the \textsc{ORAC-DR} pipeline software \citep{jeneco2015}, specifically
the \textsc{REDUCE\_SCIENCE\_NARROWLINE} recipe.
This recipe
invoked applications from the \textsc{KAPPA} \citep{currie2013},
\textsc{SMURF} \citep{chapin2013}, and \textsc{CUPID} \citep{berry2007}
packages, all from the Starlink collection \citep{currie2014}. While
the applications called by the recipe were from the Starlink 2018A
release\footnote{\url{http://starlink.eao.hawaii.edu/starlink/2018ADownload}},
the \textsc{ORAC-DR} code included improvements made since that 2018A
release\footnote{The latest code was at Commit b56f919f3f15e on Github
\url{https://github.com/starlink/ORAC-DR}.} to address specific
survey requirements.  \citet{jenness2015} provided a detailed
description of the workings of the heterodyne recipes in
\textsc{ORAC-DR} at the time of its writing.

Since the R1 data processing was finished, the reduction recipe has
undergone many improvements, yielding better-quality products.  The
highlights include flat-fielding; masking of time intervals of spectra
in Receptor\footnote{HARP terminology for a detector} H07 affected by correlated noise called ringing
\citep{jenness2015}; and automated removal of emission from the
reference (also called off-position) spectrum that appear as absorption
lines in the reduced spectra, which can bias baseline subtraction and
affect flux measurement of the emission. Since the removal of off-position
signal was not written before \citet{jenness2015}, we describe the
procedures that we adopted in Section~\ref{sec:off-pos}.

Reductions were performed at least twice, the first automated attempt
permitted visual inspection to assign recipe parameters, the most
important being the baseline and flat-field regions, whether or not
there is off-position signal or ringing spectra to remove. The recipe
parameters used to control \textsc{REDUCE\_SCIENCE\_NARROWLINE} are
available at the COHRS
repository.\footnote{\url{https://github.com/HollyThomas/COHRS/tree/master/recpars}}
An annotated example parameter file is given in Appendix~\ref{app:recpar}.

In its quality-assurance (QA) phase, the recipe used statistics to
reject outlier spectra arising from non-astronomical sources, be it
alternating bright and dark spectral channels or deviant baselines,
both transient and persistent. The former typically rejected about 0.2
per cent of COHRS spectra.  Due to ringing in Receptor H07 in about 5
per cent of observations, 5--10 percent of the spectra therein were rejected.
In about 65 per cent of cases one to three entire receptors were
expunged because of highly non-linear baselines or overwhelming
external noise sources.  A small number of observations were further
excluded because they failed to meet quality-assurance thresholds,
such as a maximum permitted $T_\mathrm{sys}$.  The QA parameters
for COHRS are listed in Appendix~\ref{app:qa}.

The second phase of the recipe converted the time-series spectra into
position-position-velocity (PPV) spectral cubes, grouping both
components of the weave pattern \citep{buckle2009} and for some
regions it incorporated data from multiple nights in order to yield a
tolerable signal-to-noise ratio. PPV cubes were re-gridded to 6-arcsec
spatial pixels, convolved with a 9-arcsec Gaussian beam, resulting in
16.6-arcsec resolution.  The released data were also re-gridded along
the velocity axis such that three raw channels became two channels,
resulting in $\Delta V = 0.635$\,km\,s\,$^{-1}$. As can be seen in
Table~\ref{tab:surveys}, this higher resolution is more comparable to
other surveys than the $\Delta V = 1.0$\,km\,s\,$^{-1}$ of R1. This improvement will make structural analyses more effective.  For example, the cloud catalog generated by \citet{colombo2019} using the \textsc{SCIMES} catalog was limited in its ability to recover low-mass molecular clouds by the coarse velocity resolution.  It also generates less channel-to-channel covariance than for a velocity width of an arbitrary round number.

In the second reduction run, only one iteration to refine baseline
regions was necessary, aided by the chosen recipe parameters. R2
retained linear baseline fitting to avoid the creation of a small
artificial dip across the emission regions excluded from
the fitting process.  Note that this contrasts with the fourth-order
polynomial adopted by CHIMPS \citep{rigby2016}.  A first-order
polynomial proved sufficient for the vast majority of COHRS 
observations.

The receptor-to-receptor responses were removed using a variant
of the \citet{curtis2010} summation method with user-defined velocity
limits in which the median intensity approximately exceeded 0.2\,K. In
under ten per cent of observations, mostly for $l>50\arcdeg$, there was
insufficient signal to determine a reliable flat-field; the errors in
the response ratios being comparable or larger than typical ratios
themselves. Normalisation was with respect to Receptor H05, except
in the 14 cases where this receptor had failed QA, whereupon the
reserve H10 became the reference receptor.  On average H10 was 1.0\%
more sensitive than H05 for the COHRS data.

Mosaics of width 5\arcdeg\ were formed from groups of PPV cubes with the
\textsc{PICARD} \citep{gibb2013} recipe \textsc{MOSAIC\_JCMT\_IMAGES} in
\textsc{ORAC-DR|} combined with task \textsc{WCSMOSAIC} from the
\textsc{KAPPA} package.  Spectral alignment was in the kinematic local
standard of rest (LSRK).\footnote{The published cubes in R1 exhibit
velocity shifts compared with those measured in the LSRK, typically one
spectral channel lower at $l=15\arcdeg$. The shift arose during mosaic
creation, because the astrometric software aligned in the heliocentric
standard of rest, despite all the input cubes from ORAC-DR being in the
LSRK.}  It allocated pixel contributions using a
Gaussian distribution with a full-width at half maximum of one pixel.
The most-central cube in each mosaic was assigned to be the reference
for the world coordinate system, so as to minimise the distortions
mapping from the galactic coordinates to a rectilinear pixel array.
Then the tiles that form R2, which abut their neighbors, were
extracted from the mosaics.

\subsubsection{Correction of off-position signal} \label{sec:off-pos}

A significant deficiency with Release 1 was the presence of absorption
features in 72 per cent of observations due to the existence of emission in the
off-position (also known as the reference) spectrum, itself still at
relatively low galactic latitudes. For Release 2 we have attempted to
remove all the detectable off-position features.

Hitherto, when circumstances make retrospective direct observations of
the off position impossible, a common approach to dealing with a
source in the reference position is to interpolate across the
locations of such absorption. However, should these interpolated
locations overlap a narrow emission line, the emission line could be
erroneously weakened, or even eradicated. Likewise if the emission is
varying rapidly downward, mere interpolation may over-compensate for
the reference signal. Since the off-position features appear in every
spectrum of an observation, we concluded that a better approach would 
be to determine the modal spectrum over regions devoid of emission or
have minimal flux, then interpolate. Further, as the receptors look at 
slightly different spatial locations, a modal spectrum should be
derived for each receptor. The difference between each interpolated modal
spectrum and its original modal spectrum derives an estimate of the
reference signal for the corresponding receptor.

In outline, the off-position contamination mitigation operated as
follows. For each receptor, the algorithm formed a pair of approximate
reference spectra. The first of the pair was derived from the
time-series cube where detected astronomical emission had been masked,
whereas the other originated from the raw time series.

At first glance, using the emission-masked data ought to be sufficient,
as it provides better discrimination between genuine dips arising from
multiple-source emission at different velocities and ones due to
off-position absorption lines. In practice, however, incomplete
emission-line detection, due to noise dominating in the line wings,
often left the off-position lines in steep-sided valleys, and as a
result were underestimated -- typically by 0.1--0.2\,K -- the depth of
these lines, thus left residual absorption lines. In contrast the
unmasked spectrum offered better estimates of the depth of reference
spectral lines. However, the unmasked modal spectrum sometimes had
difficulties discriminating between weaker source emission and a
reference line with spectrally extended emission. The aim of using
both modal spectra was to combine their assets: locate the lines with
the masked version, and determine the line strengths from the unmasked
version.  Then the algorithm refined each of the approximate reference
spectra to exclude source emission and background, to form an
estimated reference spectrum.

Since the estimated reference spectrum should have a flat baseline at
zero, accurate removal of the baseline is desirable.  In practice this
proved difficult in the presence of source emission, which might be
extended and weak over a wide velocity range. Our algorithm took an
iterative approach of twice measuring and masking lines from the
off-position then from sources. The line masking yielded better
estimates of the background signal, which after subtraction led to
better estimates of the line strengths, and an improved baseline.
The derived reference spectrum is subtracted from every spectrum in
the time-series cube. Details of the automated algorithm are provided
in Appendix~\ref{app:off-pos}.

The automated algorithm left no perceptible absorption feature for
about a third of the
lines, but over half of the lines were only partially corrected,
mostly caused by the noise raising the subtracted base level, thus not
quite removing all of the reference line.  A typical residual was
0.04\,K. The method was also less reliable for reference lines located
where there was varying and much broader source emission.  In about a
tenth of cases the reference line was not removed at all or left a
prominent line, albeit much weaker, but still could be several 
tenths of a kelvin deep in the most extreme cases.

For the cases where reference lines were still present after the
automated filtering, an additional processing step was performed. This
required the velocity limits of the residual reference lines to be
supplied via a recipe parameter. First, these line regions were masked
in the modal spectra for each receptor. Then a smooth function based
on iterative approximate solutions to Laplace's Equation filled the
gaps. The difference between the interpolated and original spectra
then yielded estimates of the residual reference line.

Despite this further attempt, off-position lines stubbornly remained
in seven observed sections of the survey. The cause was the presence
of a sheet of emission at the velocity of each absorption line, where
the sheet spanned the bulk of the spatial pixels. As a consequence,
the median or modal spectrum was representative of the emission,
rather than near or at the baseline.

To circumvent this obstacle, for each survey section we manually extracted
a polygonal spatial area devoid of emission at the line velocity. For
each area we computed the median spectrum. This was smoothed with a
Gaussian kernel of 25 channels full width at half maximum to determine
the residual baseline. The smoothing was barely affected by the
off-position lines that only spanned a few channels. Subtraction of
the smoothed spectrum from the original median spectrum resulted in a
flat baseline at zero. Although in four cases, where the off-position
line was located in strong emission, a small offset (ranging from
0.003 to 0.03\,K) correction was applied to bring the neighboring
baseline to zero. Spectral channels beyond the off-position absorption
line were set to zero. The name of this estimated reference spectrum
was passed to \textsc{REDUCE\_SCIENCE\_NARROWLINE} through a recipe
parameter, so that it could be added to the reference spectrum formed
by the previous method when the observations were reduced again.

We performed sanity checks of our off-position corrections by
comparing the median spectra of the same overlapping regions of
adjacent observations. In order to compare like with like, undefined
spectra in either observation were masked in both regions, and the
spectra were aligned along the spectral axis.  Each overlap region
typically contained 20,000 spectra.  Besides giving confidence in the 
existing corrections, these revealed many weak off-position signatures,
as evidenced by similar dips in one observation compared with all of
its neighbors.  For some overlaps, both neighbors exhibited a co-located
residual absorption feature, thus the procedure required a few iterations.
This was particularly evident for $l= 22$--33\arcdeg  where the Aquila
Rift gave rise to a common 8~$\kms$ off-position line spanning a sequence
of regions.  These were corrected in 72 cases by using the second
semi-automated method described earlier, usually with adjustments to the
previously estimated line bounds.  If that failed wholly or partially, the
manual approach was adopted in 59 cases.  For this final resort
variance-weighted average displacements from the neighboring median
spectra within the velocity limits of the off-position signal were used
to form a spectrum to be subtracted.

\section{Data Release 2} \label{sec:R2}

We provide 106~tiles in FITS format with a 0\fdg5-longitude width and a full latitude range ($0\fdg5 \times 1\fdg0$ per one tile).
All of them are perfectly contiguous.
Each tile is named in the form of central Galactic coordinate values and suffixes indicating the nature of the file,
e.g.,  COHRS\_09p50\_0p00.
Exceptionally, the first tile with the phrase `09p50' in its name is in the range of $l=$ 9\fdg5 to 9\fdg75 with a longitudinal width of 0\fdg25.
All intensities are in units of $\TAstar$.
The corrected antenna temperatures $\TAstar$ can be converted to main beam brightness temperatures $\Tmb$
by dividing $\eta_\mathrm{mb} = 0.61$.
Along the velocity axis, the data cubes in the second release 
have been cropped to a range of $-200~\kms < \vlsr < +300~\kms$,
which is extended compared with that of the first release ($-30~\kms < \vlsr < +155~\kms$).
The R2 data cubes can be obtained online from the CANFAR data archive (\dataset[doi:10.11750/22.078]{\doi{10.11570/22.0078}}).
We note that the comparison plots for R1 and R2 are also stored in the same archive.

\begin{figure}
\centering
\includegraphics[width=80mm]{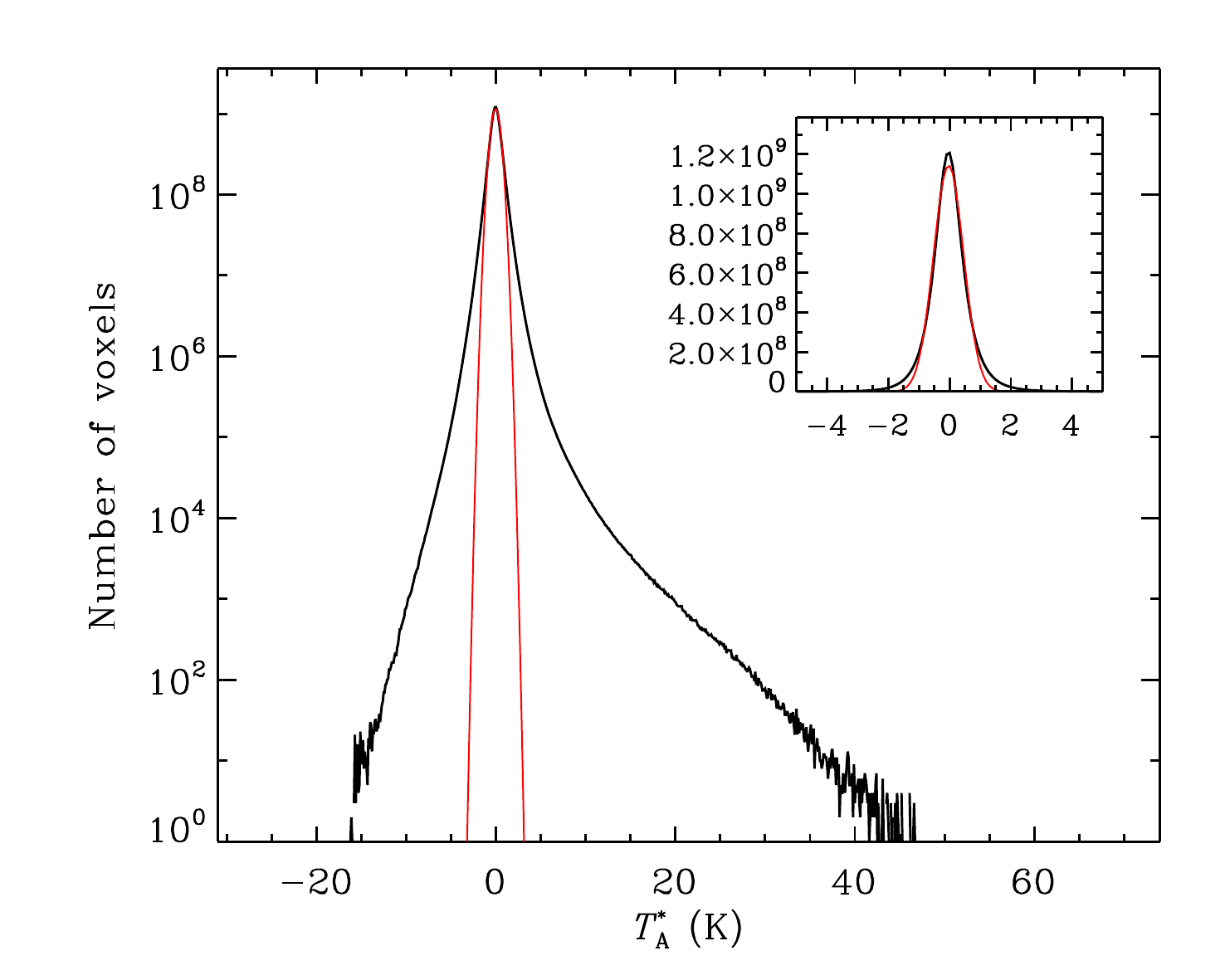}
\caption{
Histogram of the corrected antenna temperature for all voxels in COHRS, displayed on a logarithmic scale.
The bin width is 0.1\,K.
The red line shows a Gaussian fit to the distribution described by
$1.14\times10^9\,\mathrm{exp}(-\frac{1}{2}(\TAstar-0.032)^2/0.49^2)$.
The inset show the same distribution on a linear scale.
}
\label{fig:histo_voxels}
\end{figure}

Figure~\ref{fig:histo_voxels} shows a histogram of the corrected antenna temperature ($\TAstar$) for all voxels.
The values can be modelled as a normal distribution with a mean value of 0.036\,K
and a standard deviation of 0.49\,K.
The distribution shows a strong positive tail and a relatively weak negative tail.
While the former is primarily due to voxels containing \COthtotw\ emission,
the latter is significantly affected by random noise fluctuations in voxels with much higher-than-average noise levels 
(e.g., many of them at $|b| \gtrsim 0\fdg3$). 
The tile name and the mean root-mean-square (RMS) noise for each tile are tabulated in Table~\ref{tab:tiles} of Appendix~\ref{app:avg_rms}.

\section{Noise Characterization} \label{sec:noise}  

\begin{figure}
\centering
\includegraphics[width=80mm]{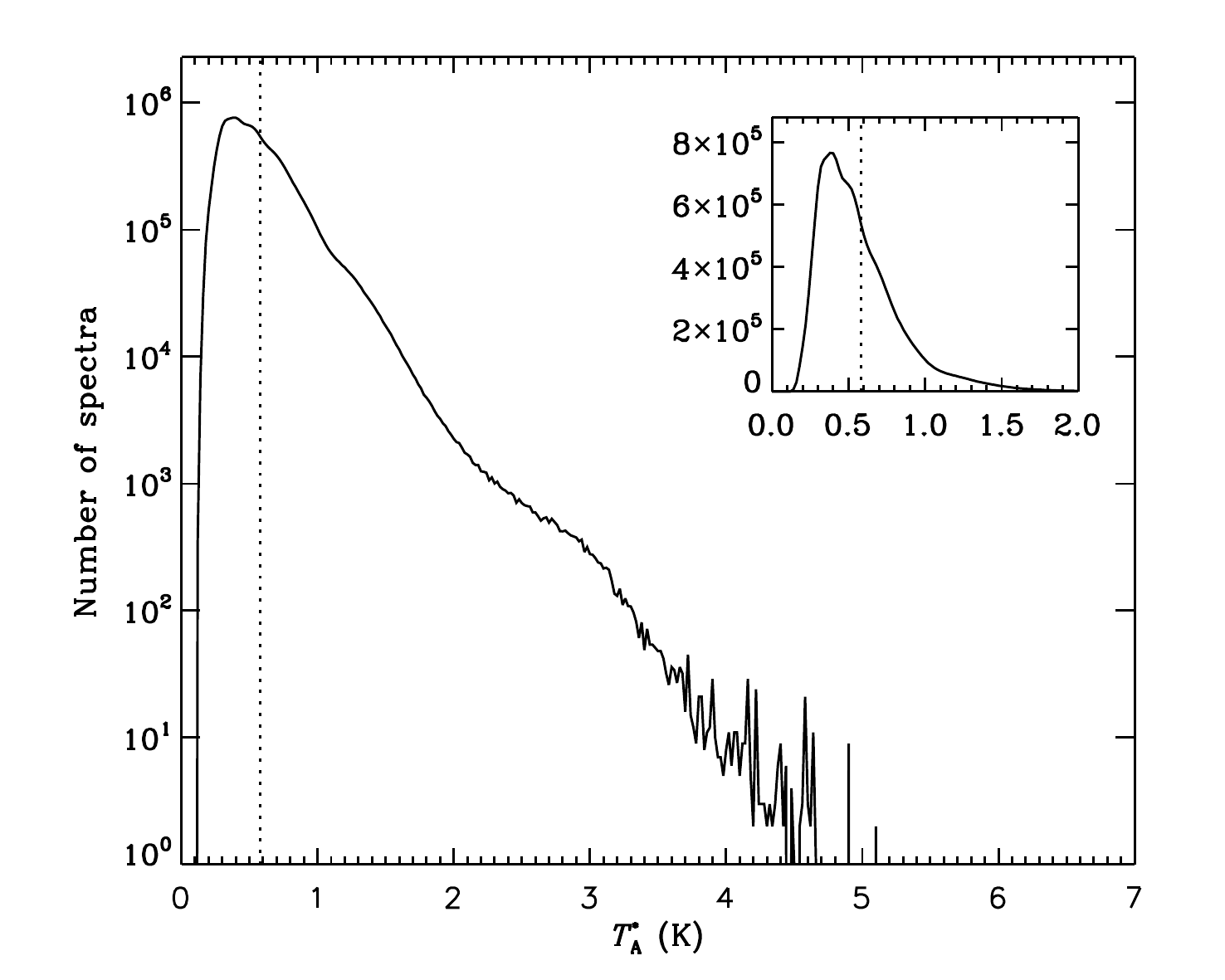}
\caption{
Histogram of the noise for all pixels in COHRS, displayed on a logarithmic scale.
The bin width is 0.02\,K.
A dotted vertical line indicates the mean value of noise levels across the survey.
The inset shows the same data on a linear scale.
}
\label{fig:histo_noise}
\end{figure}

\begin{figure*}
\centering
 \includegraphics[width=180mm]{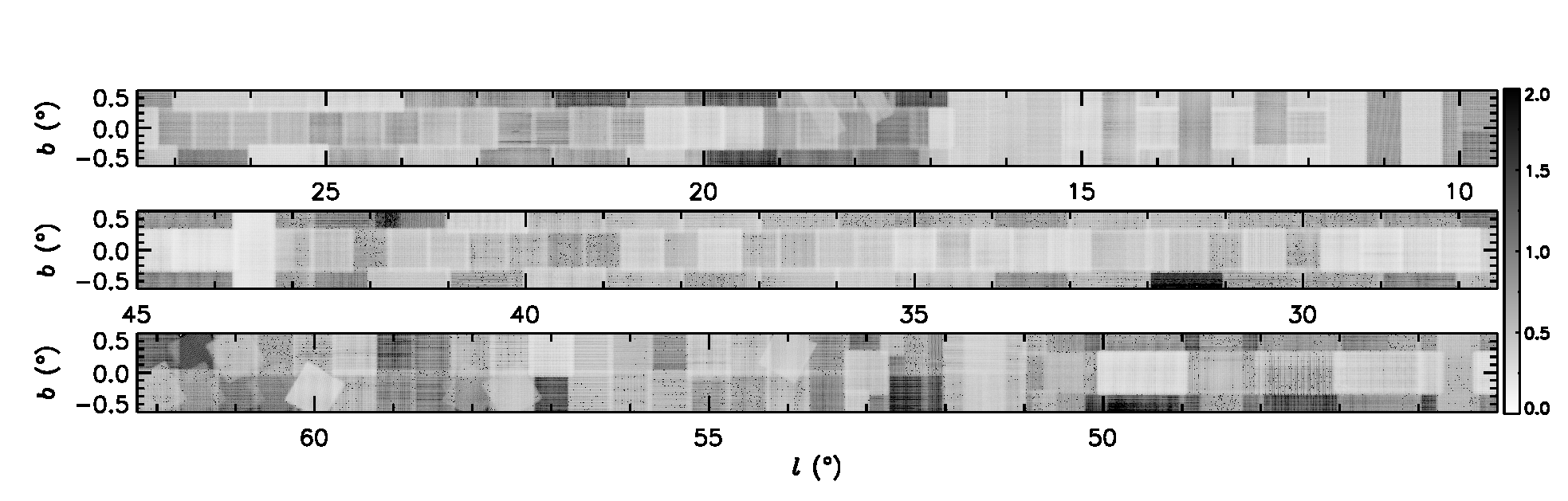}
\caption{
Noise maps for the COHRS data. 
The noise level is obtained by the conventional way that
calculates RMS values over velocity ranges where is no astronomical signal (see text for details). 
The intensity scale is in $\TAstar$ (K).
}
\label{fig:map_noise}
\end{figure*}

\begin{figure}
\centering
\includegraphics[width=80mm]{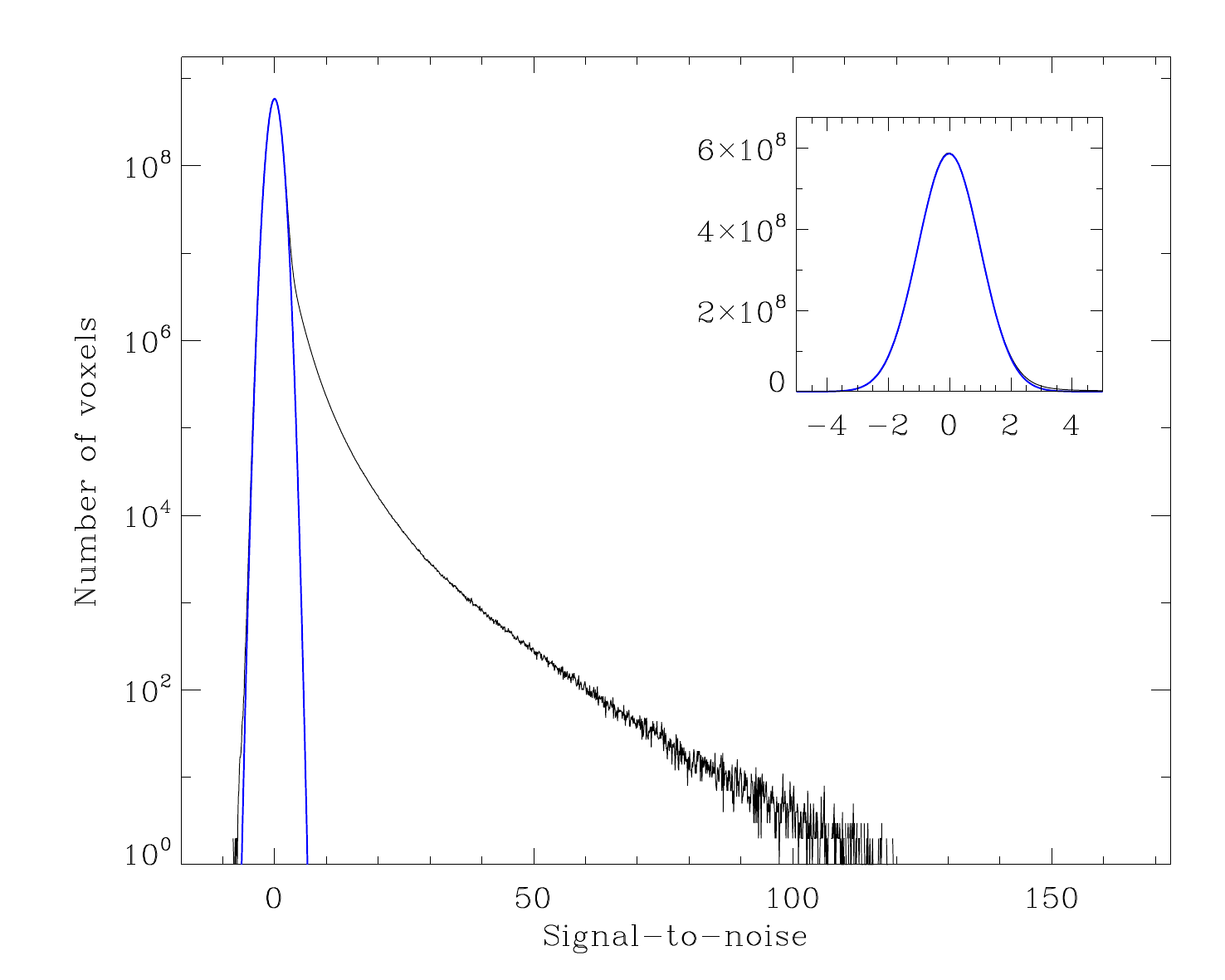}
\caption{
Histogram of the signal-to-noise ratio for all voxels in COHRS, displayed on a logarithmic scale.
The bin width is 0.1\,K.
The blue line is a Gaussian with a width (standard deviation) of 1 centered at 0.
The inset show the same distribution on a linear scale.
}
\label{fig:histo_snr}
\end{figure}

A histogram and two-dimensional map of RMS noise levels of the spectra across the COHRS areas
are shown in Figures~\ref{fig:histo_noise} and \ref{fig:map_noise}, respectively.
The noise levels were measured by taking the standard deviation of a baseline over a velocity range,  $\vlsr < -100~\kms$ or $\vlsr > +200~\kms$,
in which no astronomical signal is visible.
The histogram peaks at about 0.3\,K, close to the standard deviation of the normal distribution of all voxels shown in Figure~\ref{fig:histo_voxels},
and contains a fat tail, mainly contributed from noisiest areas at relatively high latitudes (see Figure~\ref{fig:map_noise}).
The variation across the noise map results from the combined effect of weather conditions, observing elevations, and the number of HARP receptors included in the reduction.
We found that 27\% of the pixels have RMS noise levels $<$ 0.4~K, 60\% $<$ 0.6~K, and 80\% $<$ 0.8~K.
The mean and median values of noise levels are 0.6\,K and 0.5\,K, respectively.
Figure~\ref{fig:histo_snr} shows a histogram of the signal-to-noise ratio for all voxels 
indicating the noise is Gaussian (normal distribution).
The positive tails in the log plot are due to real astronomical signals.

The average of the RMS noise values of all pixels for each tile
is given in Table~\ref{tab:tiles} of Appendix~\ref{app:avg_rms}.
For tiles with the full latitude range of R1 shown in Table 3 of \dempseypaper, the RMS noise level of those tiles in R2 is reduced to 13--55\% (36\% on average).
As mentioned in \dempseypaper, some of the final tiles contain maps observed on different nights and under different conditions, resulting in some significant changes in RMS noise levels in one tile. 
Therefore, in such a case, it is not appropriate to say that the average RMS is a representative value for an individual map.
Instead, Figure~\ref{fig:map_noise} is helpful for visualizing the spatial noise distribution throughout the survey and within each tile.

\section{Results} \label{sec:results}

\subsection{Integrated position-position maps} \label{sec:ppmaps}

Figure~\ref{fig:lbitg_vr} shows channel-maps of \COthtotw\ emission with the velocity interval of $20~\kms$
from $\vlsr= -50$ to $150~\kms$ but also an integrated map over the whole velocity range.

Most of \COthtotw\ emission in the ranges of $l \simeq 10\arcdeg$ to 22\arcdeg\ and 38\arcdeg\ to 45\arcdeg\ 
appears in negative latitudes.
Such a tendency between $l = 12\arcdeg$ and 22\arcdeg\ was reported by
\citet{umemoto2017} using FUGIN $\mathrm{^{12}CO}/\mathrm{^{13}CO}/\mathrm{C^{18}O}$ (1-0) maps.
They explained that this is a distance effect that occurs because the Sun is not located at the true Galactic midplane.
The Sun is located slightly above the true Galactic midplane, 
and an estimated offset is $\sim 10$--30~pc \citep[e.g.,][and references therein]{anderson2019, karim2017}. 
Many of the previous studies to estimate the Sun's height used the star-counts methods.
Recent studies by \citet{su2016, su2019} used a large-scale CO gas distribution
to derive the position of the Sun, yielding $\sim 17$~pc, 
which is similar to the median of the published estimates.
For objects located at the true Galactic midplane,
those closer to the Sun will appear at higher negative latitudes
while those at a large distance will converge to $b=0\arcdeg$.
On the other hand, the remaining \COthtotw\ emission regions are roughly distributed around $b=0\arcdeg$.
But we note that positive velocity channels contain \COthtotw\ components at two distances along line-of-sight, the near and far.
Therefore, \COthtotw\ emissions originating from two very difference distance ranges are stacked up, and the distance effect of this discrepancy for each line-of-sight should be carefully considered.

Many prominent bright regions appear across the COHRS area
while significant faint extended emission is detected.
Some of the bright regions include well known star-forming regions, 
such as W31 ($l= 10\fdg3$), W33 (12\fdg8), W42 (25\fdg3), W43 (30\fdg8), 
W47 (37\fdg6), W49A (43\fdg2), and W51 (49\fdg4). 
Among them, three massive and luminous star-forming regions, W43, W49A, and W51,
are presented in Section~\ref{sec:example}.

\begin{figure*}
\centering
\includegraphics[width=160mm]{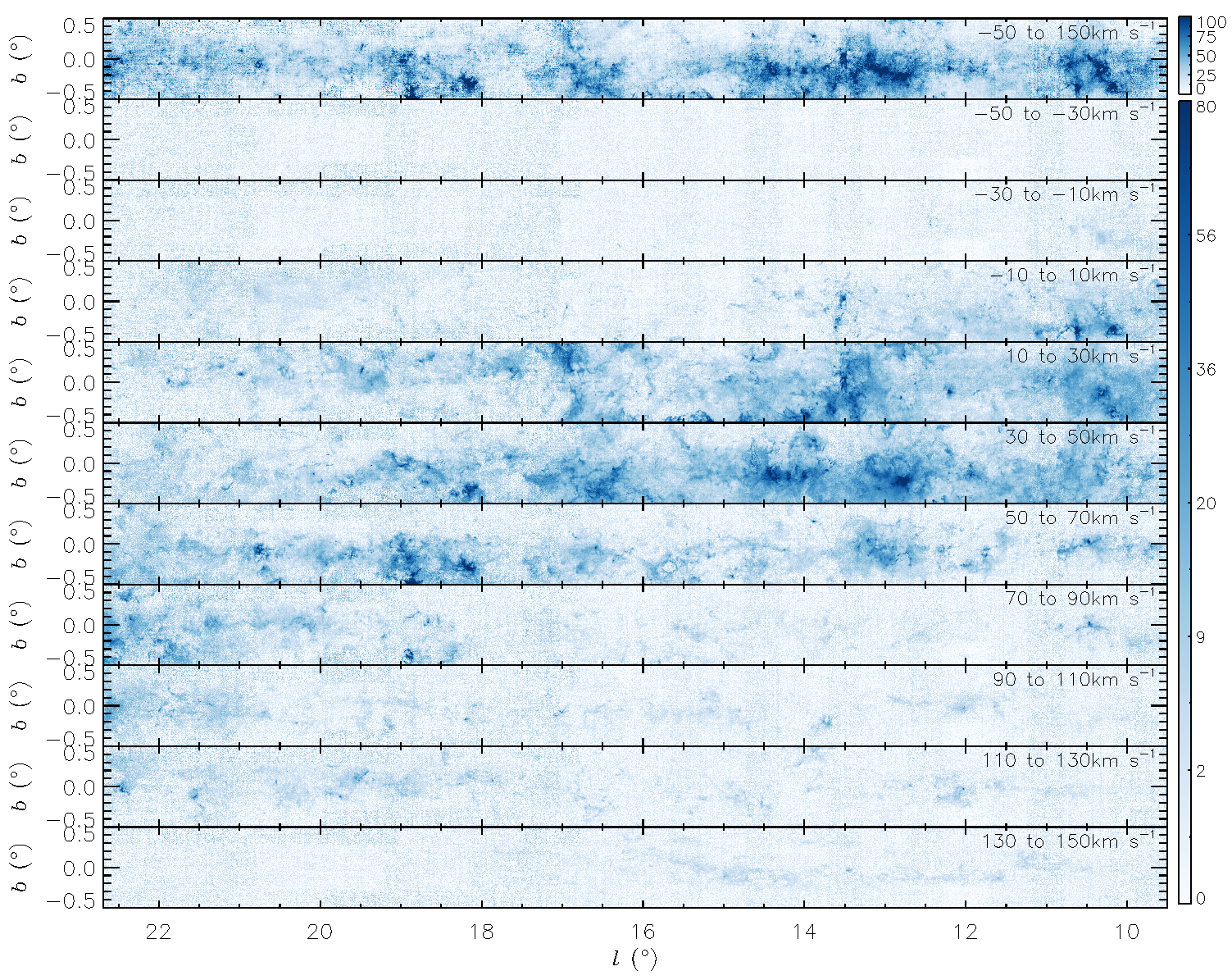}
\caption{
The maps of velocity-integrated emission ($\TAstar$) in COHRS.
This map is obtained by integrating over the velocity range written at the top of each panel. 
The units on the intensity scale are $\Kkms$. 
}
\label{fig:lbitg_vr}
\end{figure*}
\addtocounter{figure}{-1}

\begin{figure*}
\centering
\includegraphics[width=160mm]{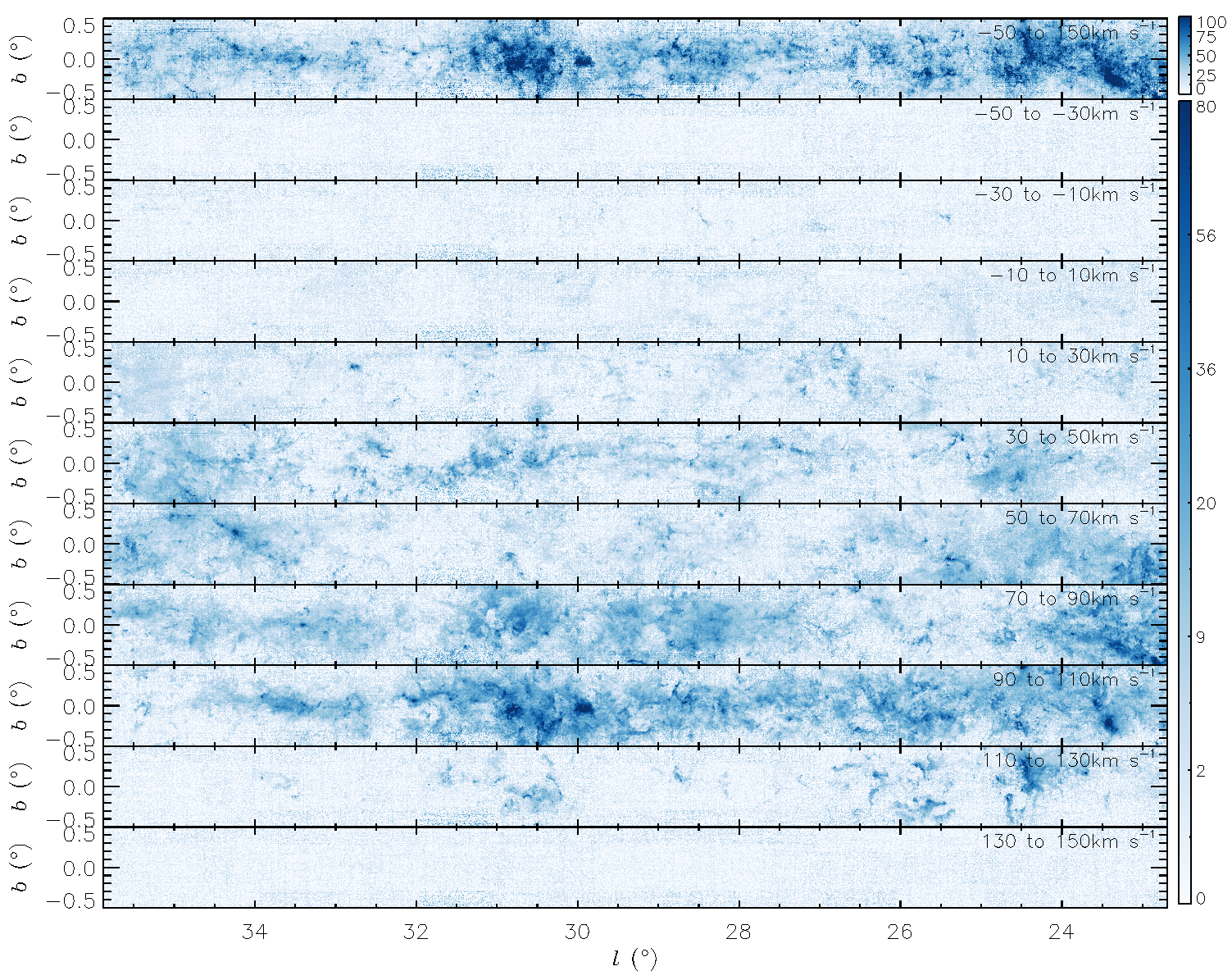}
\caption{Continued}
\end{figure*}
\addtocounter{figure}{-1}

\begin{figure*}
\centering
\includegraphics[width=160mm]{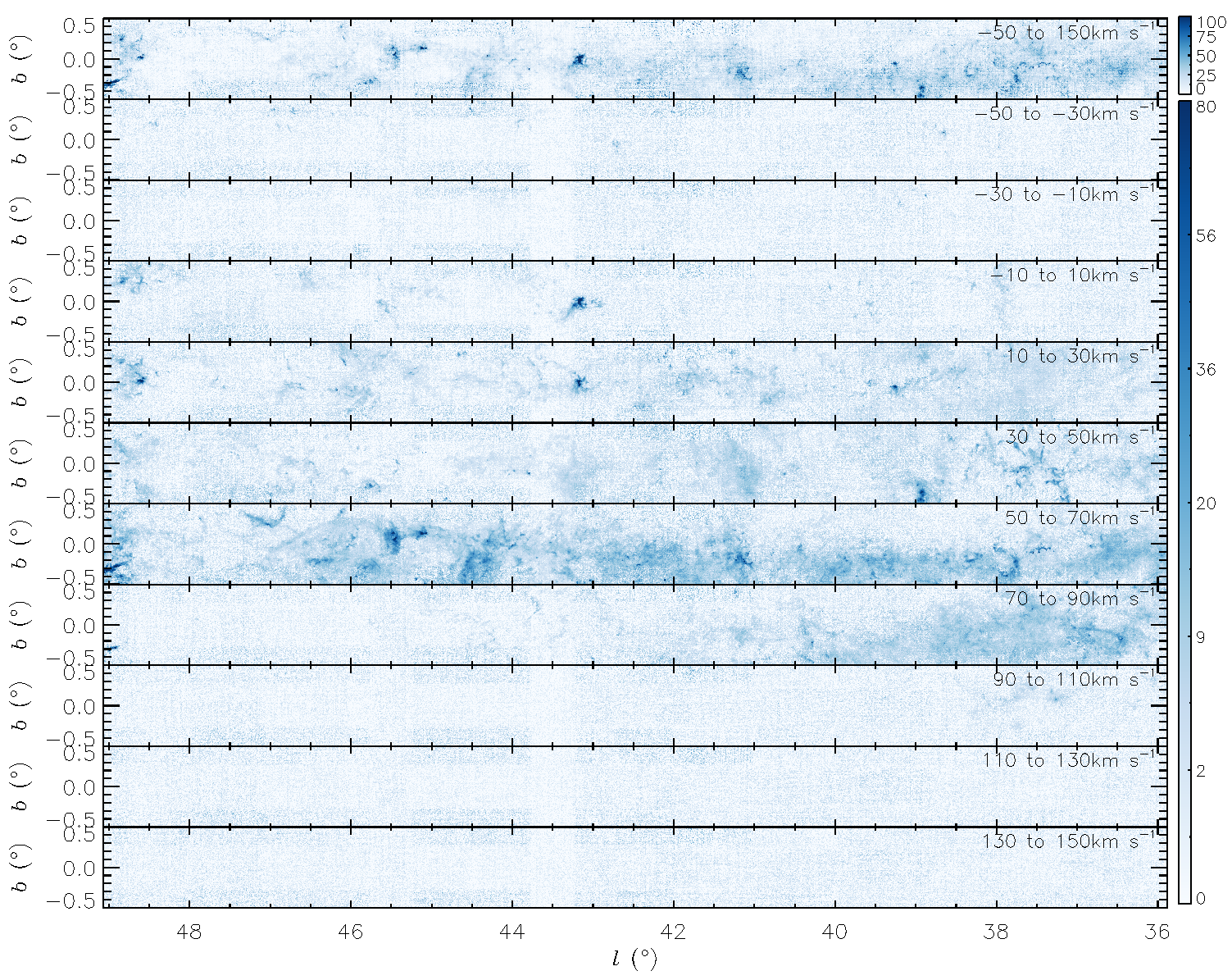}
\caption{Continued}
\end{figure*}
\addtocounter{figure}{-1}

\begin{figure*}
\centering
\includegraphics[width=160mm]{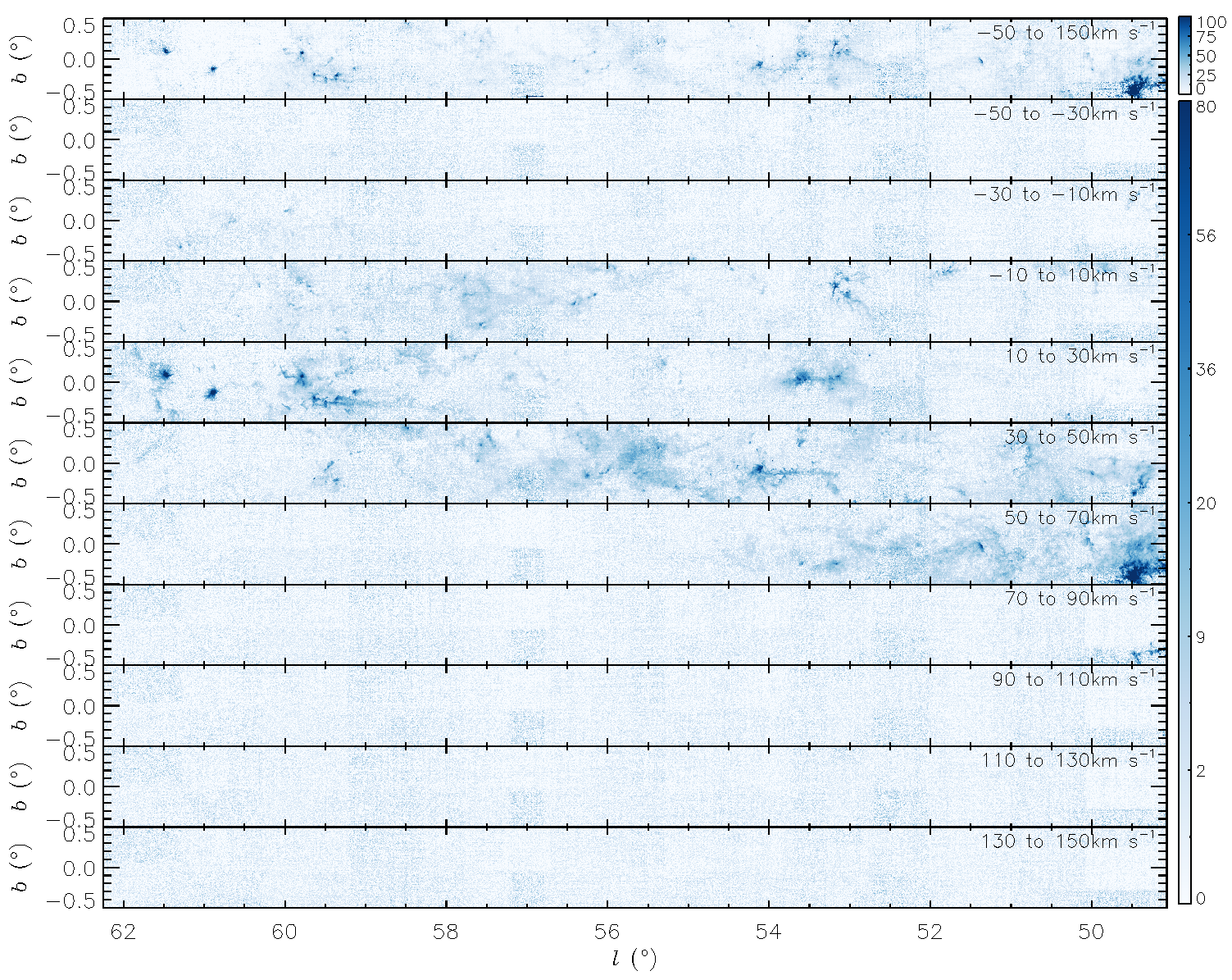}
\caption{Continued}
\end{figure*}

\subsection{Integrated position-velocity map} \label{sec:pvmaps}

\begin{figure*}
\centering
\includegraphics[width=170mm]{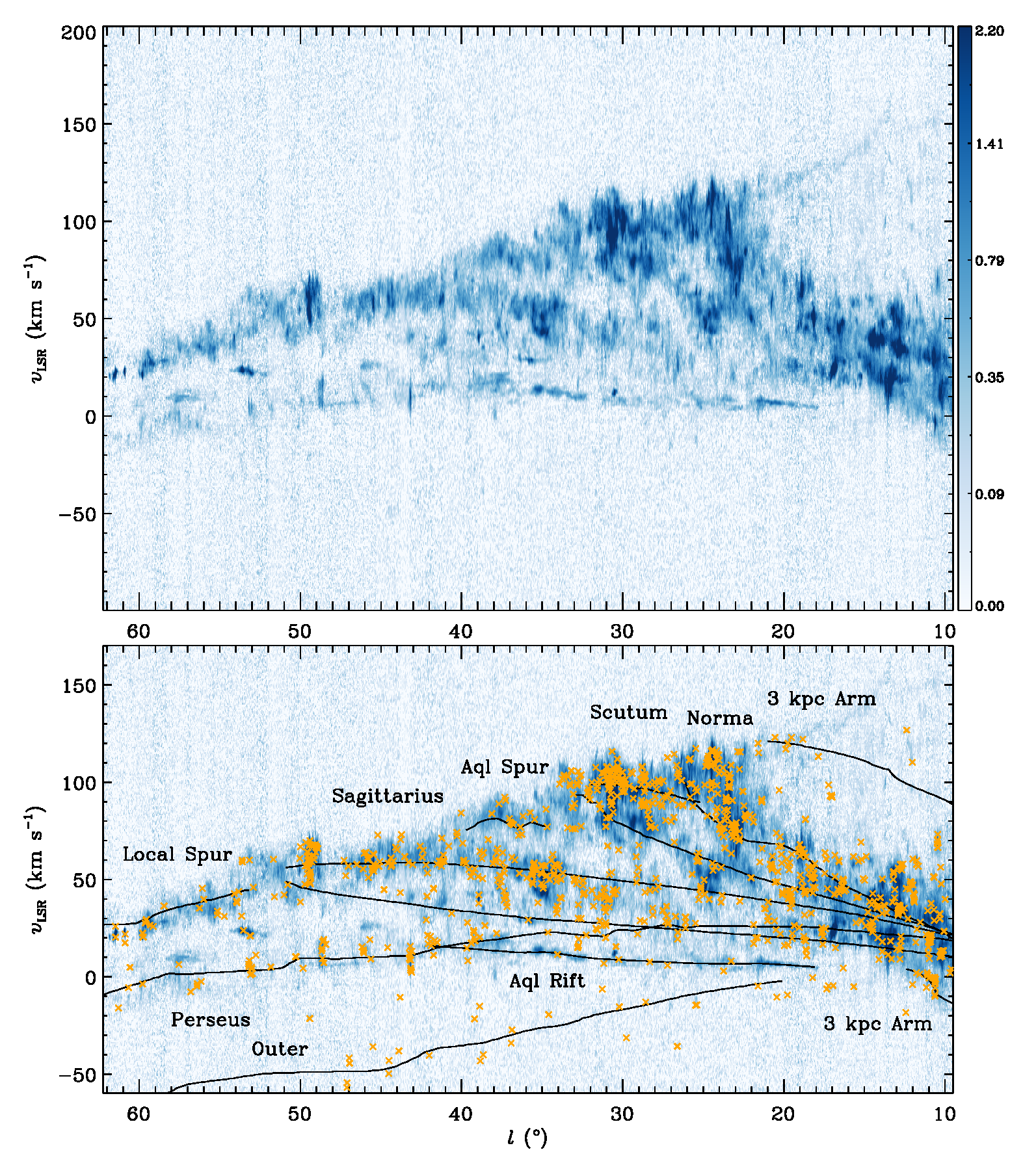}
\caption{
Position-velocity ($l-\vlsr$) map for the \COthtotw\ emission ($\Tmb$) in COHRS.
This map is obtained by integrating over the latitude axis. 
The map is drawn on a square-root scale.
The units on the intensity scale are K~degrees.
The bottom image is the same as the top, but is shown for a narrower velocity range overlaid with known \schii\ regions and spiral-arm loci.
Cross symbols indicate \wise\ \schii\ regions.
The traces of main spiral arms (Scutum, Sagittarius, Perseus, and Norma-Outer arms)
and interarm features (Local Spur, Aquila Spur, Aquila Rift, and 3~kpc arm) from \citet{reid2016, reid2019} 
are overlaid using black curves.
}
\label{fig:lvitg}
\end{figure*}

\begin{figure}
\centering
\includegraphics[width=80mm]{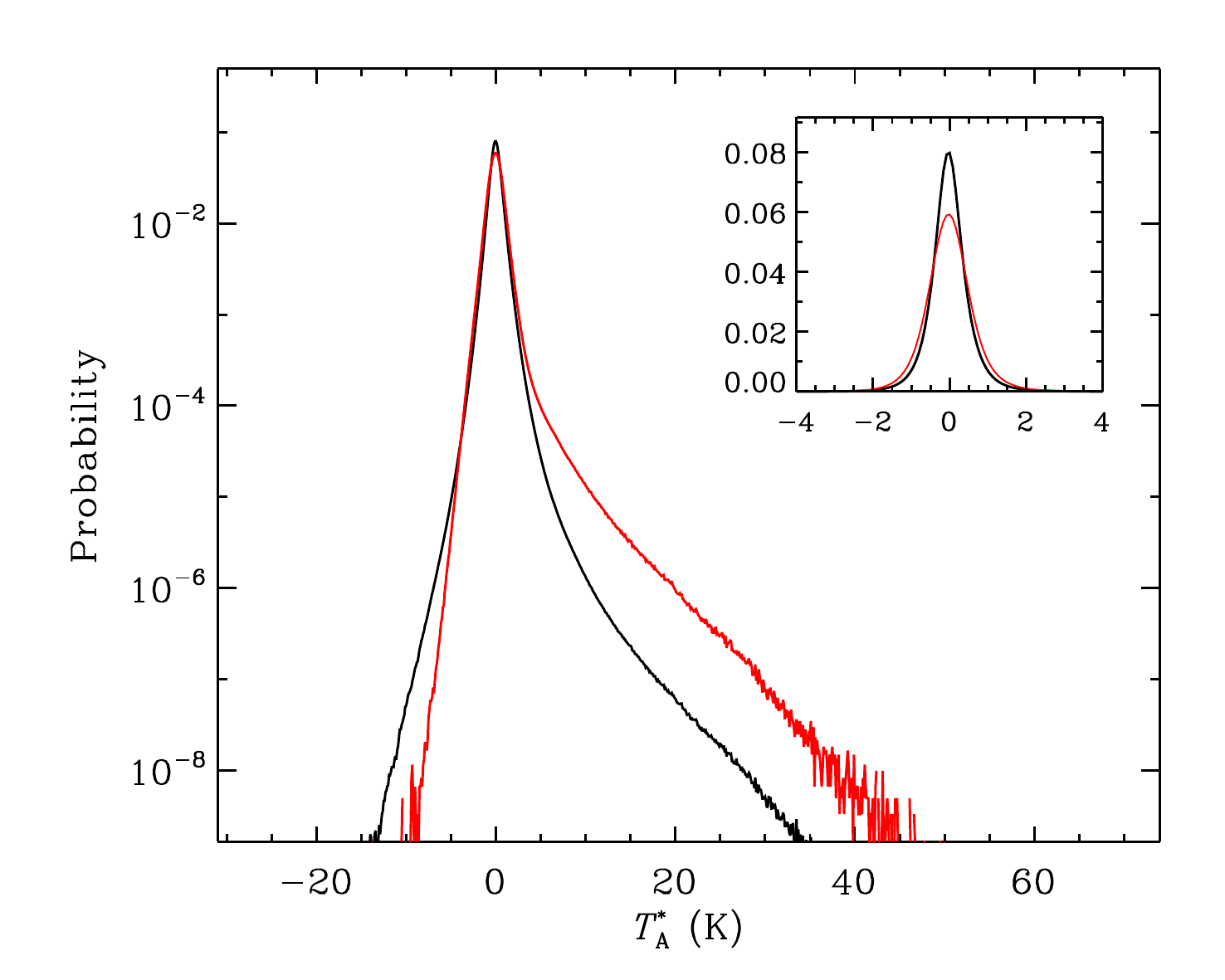}
\caption{
Temperature PDFs of all \COthtotw\ distribution (black) and the \schii\ region-related \COthtotw\ distribution (red) in COHRS. 
}
\label{fig:pdf}
\end{figure}

A position-velocity ($l$-$\vlsr$) map of \COthtotw\ emission 
integrated over the whole latitude range is shown in the top panel of Figure~\ref{fig:lvitg}.
The bottom panel presents the same \COthtotw\ $l$-$\vlsr$ map,
but known \schii\ regions and spiral arm loci overlaid.
\schii\ regions are obtained from the catalog (V2.2; \dataset[doi:10.26131/IRSA146]{\doi{10.26131/IRSA146}}) 
of the all-sky {\it Wide-Field Infrared Survey Explorer} \citep[\wise;][]{anderson2014}.
Note that \schii\ regions shown here are those given a single measured velocity,
but the original \wise\ catalog lists numerous \schii\ regions with no available velocity measurement
or many \schii\ regions with multiple velocities measured. 
The traces of spiral arms are derived from \citet{reid2016} and updated in \citet{reid2019}.
In the figure, main spiral arms (Scutum, Sagittarius, Perseus, and Norma-Outer arm)
and interarm features (Local Spur, Aquila Spur, Aquila Rift, and 3~kpc arm) are drawn.
\COthtotw\ emission, in general, agrees well with the spiral-arm traces
although those associated with the Outer arm are extremely faint and sparsely distributed.
The Aquila Rift ranges from about 17\arcdeg--43\arcdeg, from which
bright parts near $l=$ 18\arcdeg--22\arcdeg\ and 32\arcdeg--36\arcdeg\ are clearly seen.
Also, in the remaining sections, weak \COthtotw\ emission appears along where other CO isotopologue emissions are detected (cf. Figure~5 of \citealt{dame1985} and Figure~3 of \citealt{jackson2006}).
Well-studied star-forming regions such as W43 ($l$, $\vlsr =$ 30\fdg9, 95~$\kms$) and W51 (49\fdg4, 60~$\kms$)
show highly peaked emission with complex velocity structures.
The locations of \schii\ regions generally coincide with CO-bright regions. 
This characteristic is illustrated by comparing the probability distribution functions (PDFs) for the entire \COthtotw\ distribution and the \COthtotw\ distribution related to the \schii\ regions (see Figure~\ref{fig:pdf}). 
We extract the \schii\ region-related \COthtotw\ from the ($d_\textsc{hii}$, $d_\textsc{hii}$, 15) pixel bin at each \schii\ region location. 
$d_\textsc{hii}$ is the diameter of the \schii\ region given in the \wise\ catalog, and 15 pixels at velocity are arbitrarily chosen to have about 10~$\kms$. 
The PDF of \schii\ region-related \COthtotw\ has a lower peak, which is mainly contributed by noise, and a fatter positive tail, which indicates stronger CO signals, than the PDF of all \COthtotw.

\section{Example COHRS Data} \label{sec:example}

\begin{figure*}
\centering
\includegraphics[width=180mm]{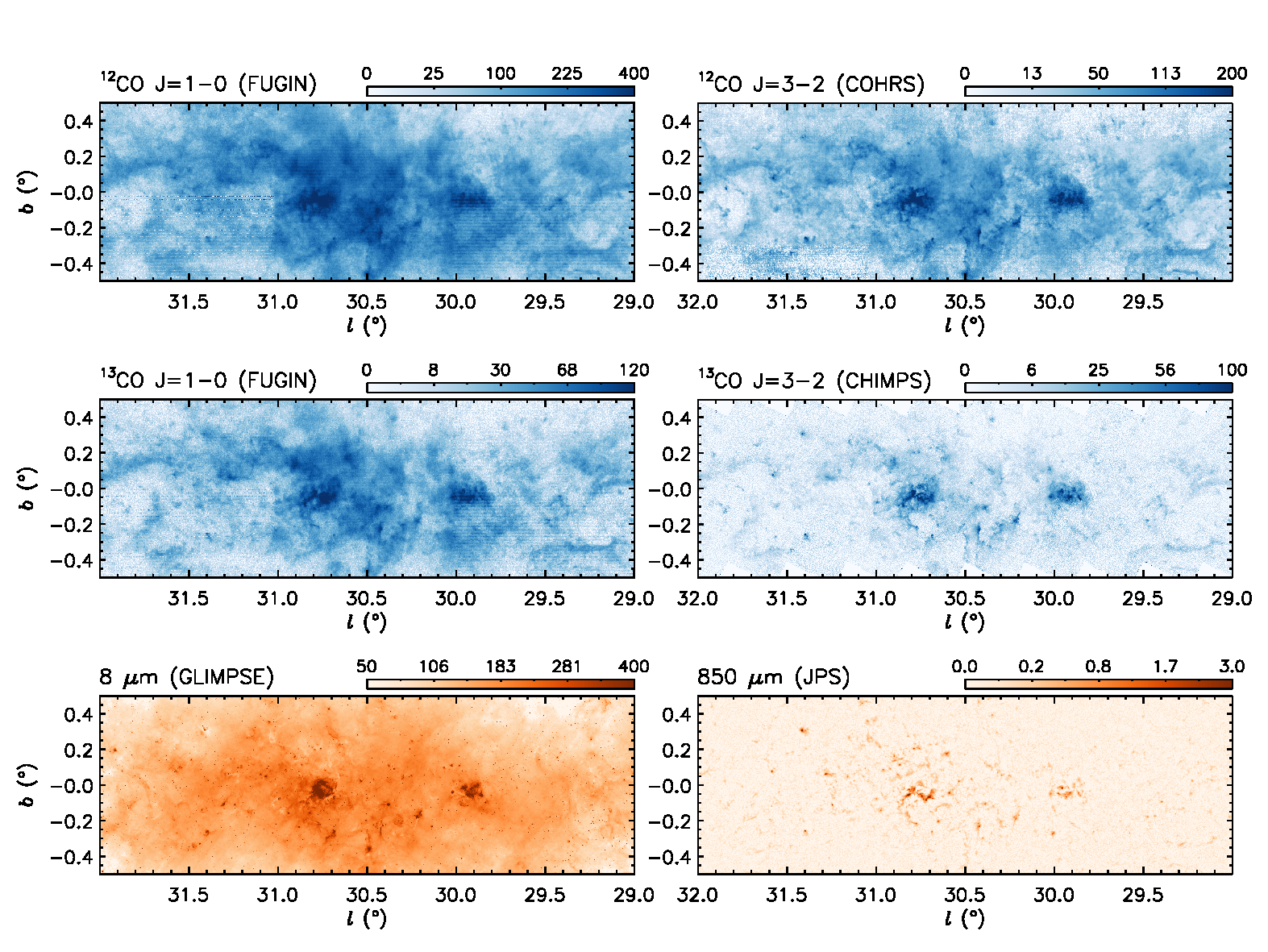}
\includegraphics[width=180mm]{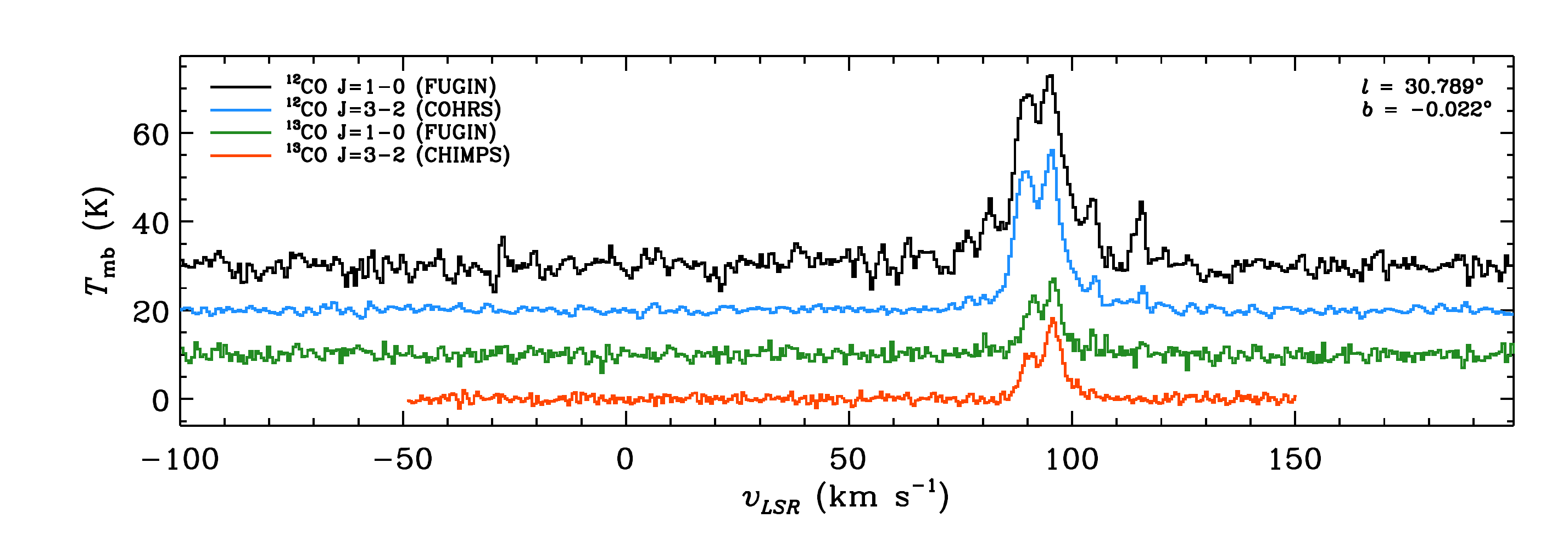}
\caption{W43 star-forming region within the COHRS survey area.
From the top left to the bottom right, each panel displays
\COontoze\ FUGIN data, \COthtotw\ COHRS data,
$\mathrm{^{13}CO}$ (1-0) FUGIN data, $\mathrm{^{13}CO}$ (3-2) CHIMPS data,
8~\um\ GLIMPSE data, 850~\um\ JPS data, and examples of CO spectra, respectively.
The CO maps are velocity-integrated over the $\vlsr$ range of (80, 110)~$\kms$ \citep{nguyen2011},
and the units on the intensity scale of the integrated main beam temperature are $\Kkms$.
The units on the intensity scale of the GLIMPSE and JPS data are \textrm{MJy/sr} and \textrm{Jy/beam}, respectively. 
The CO spectra are obtained at the position closest to the velocity-integrated \COthtotw\ emission peak.
The offsets of 10K, 20K, and 30K to the spectra have been added for better visualization.
}
\label{fig:ppm_W43}
\end{figure*}

\begin{figure*}
\centering
\includegraphics[width=150mm]{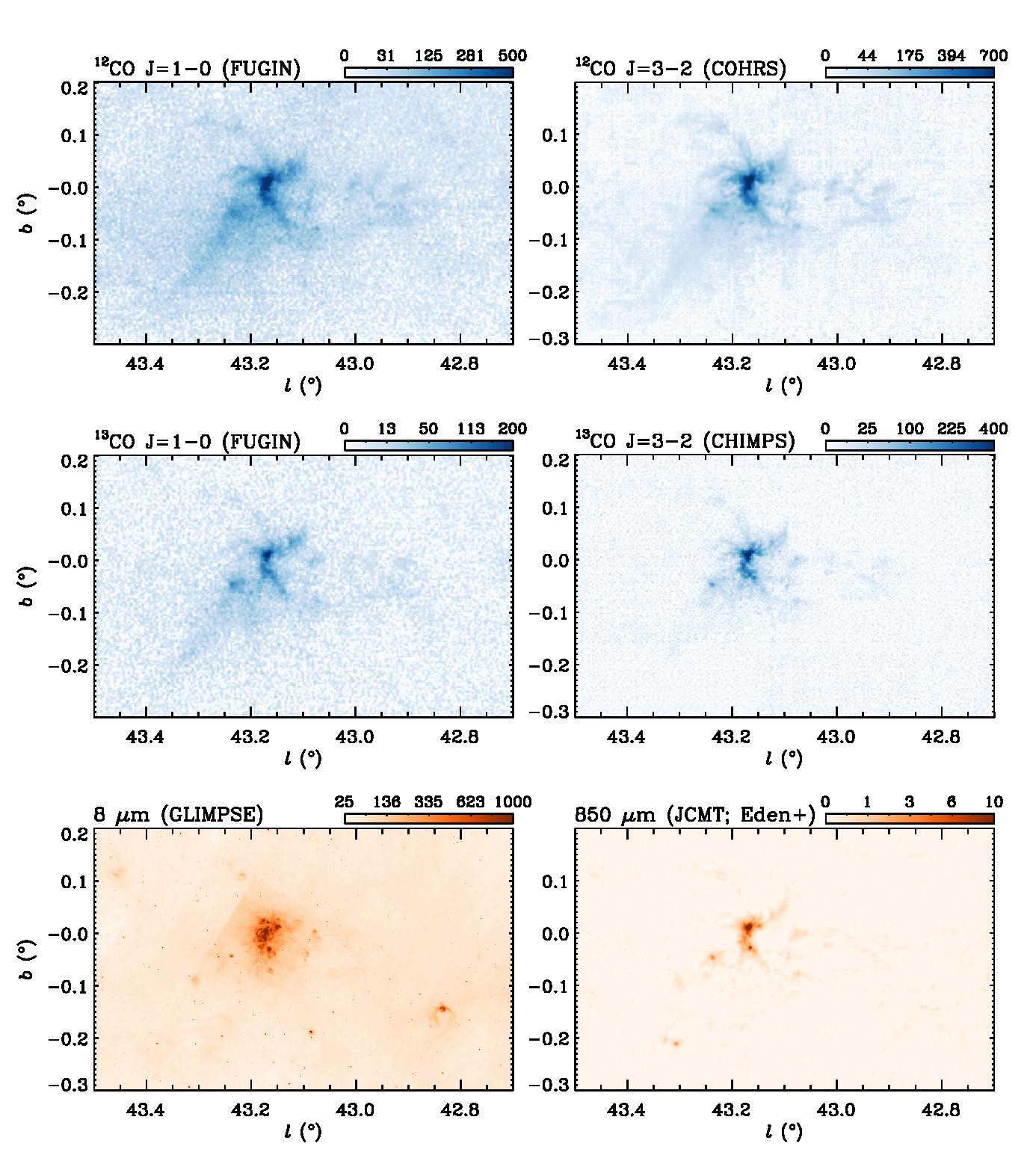}
\includegraphics[width=150mm]{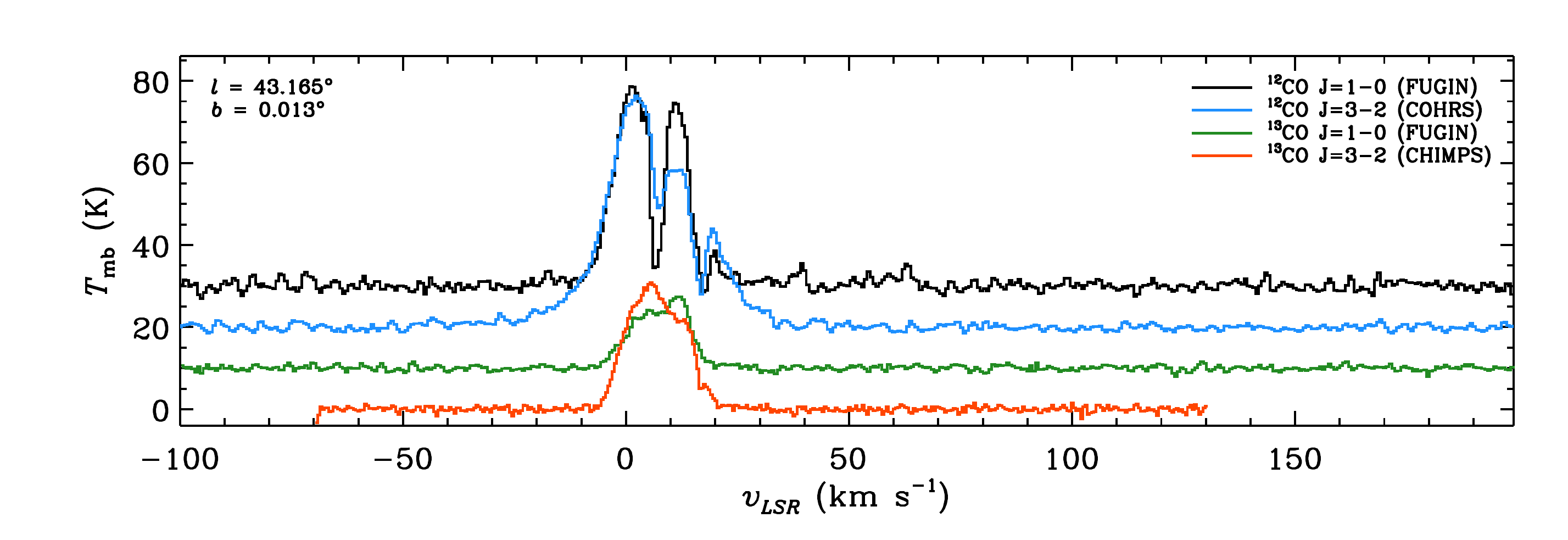}
\caption{Same as Figure~\ref{fig:ppm_W43}, but for W49A star-forming region.
 850~\um\ continuum data are taken from \citet{eden2018}.
The CO maps are velocity-integrated over the $\vlsr$ range of ($-20$, $+30$)~$\kms$ \citep{galvan2013}.
In the GLIMPSE 8~\um\ image, instrumental artifacts remain around the bright W49 area at $(l, b) \sim$ (43\fdg17, -0\fdg01).
}
\label{fig:ppm_W49A}
\end{figure*}

\begin{figure*}
\includegraphics[width=150mm]{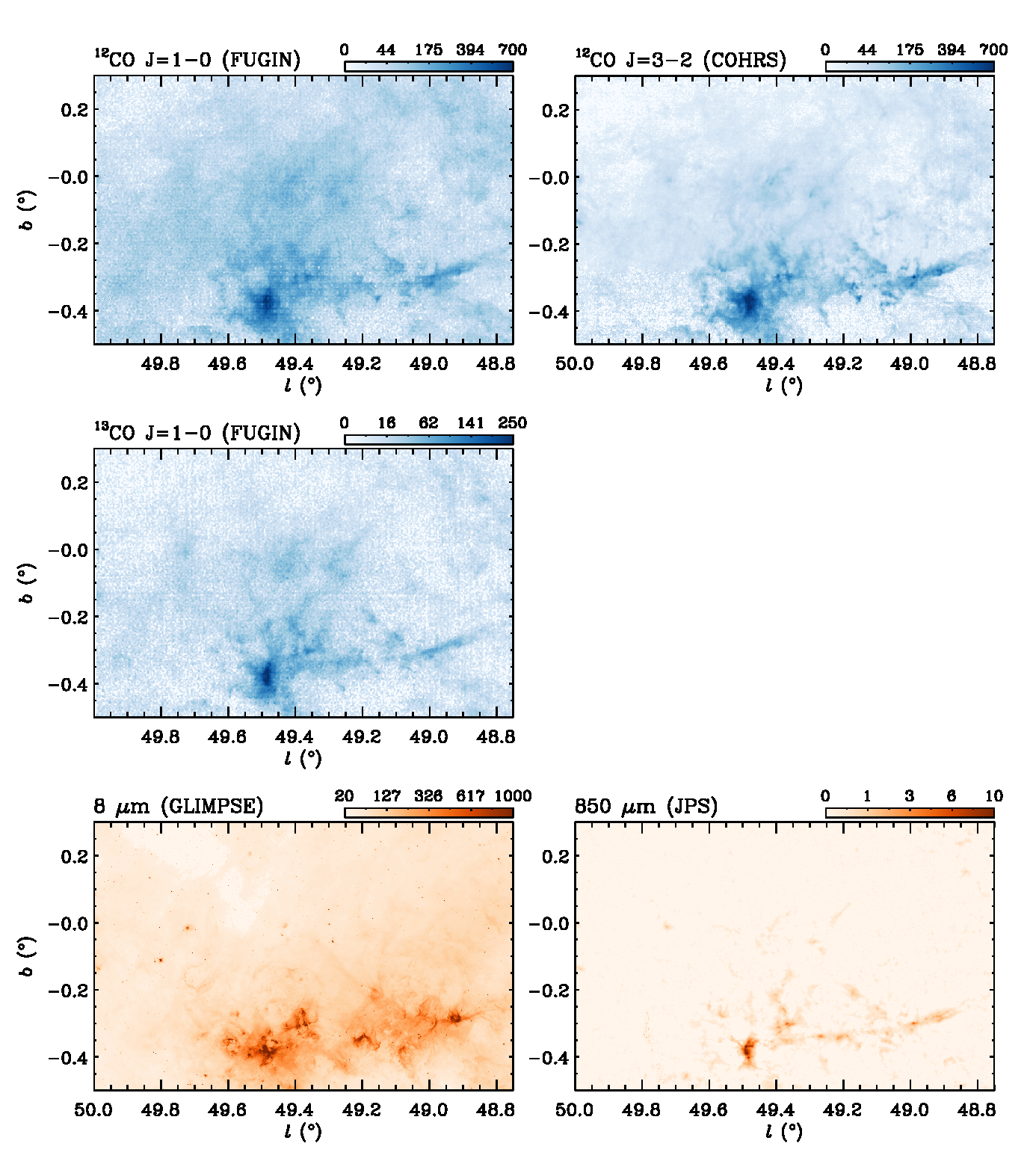}
\includegraphics[width=150mm]{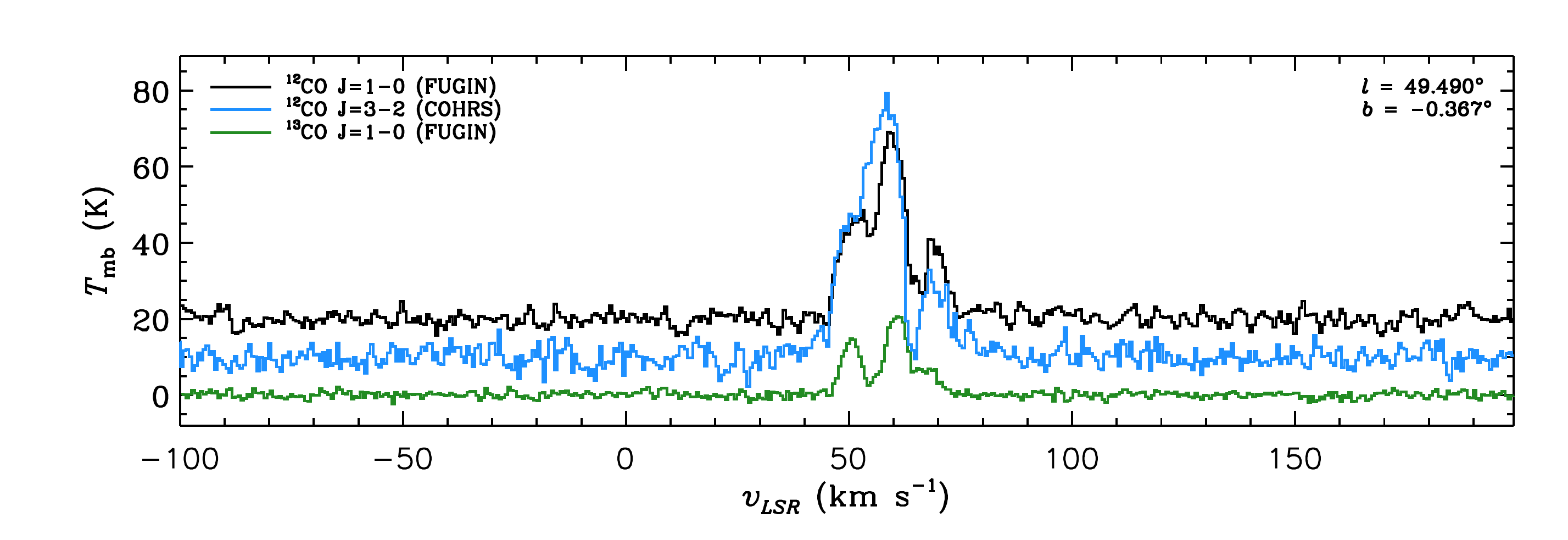}
\caption{Same as Figure~\ref{fig:ppm_W43}, but for W51 star-forming region.
The CO maps are velocity-integrated over the $\vlsr$ range of (30, 85)~$\kms$ \citep{kang2010}.
There are no available CHIMPS data for this region.
}
\label{fig:ppm_W51}
\end{figure*}

Examples of COHRS data for active star-forming regions,
such as W43, W49A, and W51, are displayed in Figures~\ref{fig:ppm_W43}--\ref{fig:ppm_W51}. 
These figures present integrated-intensity maps from two rotational transitions of two CO isotopologues
($\mathrm{^{12}CO}/\mathrm{^{13}CO}$ (1-0) and (3-2))
from FUGIN, COHRS, and CHIMPS data
together with 8~\um\ (GLIMPSE;\dataset[doi:10.26131/IRSA210]{\doi{10.26131/IRSA210}}) and 850~\um\ (JPS or \citet{eden2018}) continuum maps, if available. 
Also, examples of available CO spectral lines obtained from the nearest position to the velocity-integrated \COthtotw\ emission peak are shown.
That is, spectra of different CO isotopologues and transitions are extracted from the matching positions within the one-pixel size of each survey.
COHRS emission maps show well clumpy or filamentary structures and in addition extended diffuse CO gas compared with other CO maps. 
Also, the bright clumpy or filamentary features appear to be closely associated with star-forming regions seen in the continuum maps.

W43 is one of the most massive molecular complexes 
(total gas mass ($\Mgas$) of $\sim 7.1 \times 10^7~\Msol$, bolometric luminosity ($\Lbol$) of $\sim 8.5 \times 10^5~\Lsol$, \citealp{nguyen2011, urquhart2014b}) in the Galaxy.
The distance of W43 is estimated to be 5.5~kpc from the Sun \citep{zhang2014}.
The region is located near the tangential point of the Scutum arm, where the spiral arm meets the Galactic bar \citep{nguyen2011}.
It appears between $l=$ 29\arcdeg\ and 32\arcdeg\ and $b=$ $-1\arcdeg$ and $+1\arcdeg$
in the velocity range $\vlsr=$ 80--110~$\kms$ \citep[e.g.,][]{nguyen2011, kohno2020}.
W43 extends beyond the latitude coverage of the COHRS survey,
but the brightest parts such as W43-main ($l \sim 30\fdg8$) and W43-south ($l \sim 29\fdg9$),
which are very active star-forming regions, are covered, as seen in Figure~\ref{fig:ppm_W43}.
W43-main is considered as a mini-starburst region,
which contains fifty-one protocluster candidates \citep{motte2003}.
\citet{kohno2020} suggests that a supersonic cloud-cloud collision causes
the local mini-starbursts in W43.

W49A is another one of the most well-known massive and luminous Galactic star-forming complexes
($\Mgas \sim 1.1 \times 10^6~\Msol$, $\Lbol \sim 3.63 \times 10^6~\Lsol$, \citealp{galvan2013, urquhart2014b}), 
despite being located at a large distance of 11.1~kpc from the Sun \citep{zhang2013}.
It is centered at $(l, b) =$ $(43\fdg15, -0\fdg01)$ and
appears in the LSR velocity range from $-20$ to $+30\,\kms$.
The region is lying on the Perseus arm in the inner Galaxy.
The giant molecular cloud (GMC) of W49A extends more than 100~pc ($\sim30\arcmin$) in longitudes \citep{simon2001}, 
while active star formation occurs mainly in the central area within $\sim20$~pc ($\sim6\arcmin$) \citep{welch1987, alves2003}. 
As shown in Figure~\ref{fig:ppm_W49A}, 
the bright CO emission region is aligned along the line of sight with the central star-forming region, 
and CO maps show clear hub-filament structures \citep{galvan2013}.
A ring or shell-shaped, faint CO emission feature extends westward from the central bright region.
The extended feature is visible in \COontoze, but clearer in \COthtotw.
Much weaker and clumpier $\mathrm{^{13}CO}$ emission features are discernible where the structure shown in the \CO\ maps is located.
The \CO\ profiles in Figure 8 show a double peak feature due to self-absorption at $\vlsr$ $\sim 7\,\kms$, which is stronger in the (1-0) line than in the (3-2). 
On the other hand,  the two transitions of $\mathrm{^{13}CO}$ have different velocity profiles, indicating that the two transitions trace different internal conditions.

W51 is also a particularly prominent massive and luminous Galactic star-forming complex
($\Mgas \sim 1.2 \times 10^6~\Msol$, $\Lbol \sim 4.68 \times 10^6~\Lsol$, \citealp{carpenter1998, urquhart2014b}).
It is estimated to be at a distance of 5.4~kpc from the Sun \citep{sato2010}
and located near the tangent point of the Sagittarius arm.
The region is centered on $(l, b) \approx$ $(49\fdg4, -0\fdg3)$,
and appears as a long filamentary stream with a length of $\sim 100$~pc,
which is mostly covered by COHRS data (see Figure~\ref{fig:ppm_W51}).
W51 GMC is distributed in a broad velocity range of $\vlsr =$ 30--85~$\kms$ \citep{kang2010},
and embeds two star-forming regions, W51A and W51B, and host a supernova remnant W51C.
The brightest CO-emission region near $(l,b) \approx$ $(49\fdg5, -0\fdg4)$ shown in Figure~\ref{fig:ppm_W51}
is associated with W51A.
$\mathrm{^{12}CO}$ profiles show a clear double peak due to self-absorption around $\vlsr =$ 65~$\kms$.
The self-absorption feature is stronger in the (3-2) compared with the (1-0), 
indicating foreground gas is likely colder than the gas associated with the W51 complex.
However, the self-absorption situation can occur in subthermal excitation, where the density is smaller than the effective critical density, even if the gas is not colder. Therefore, additional analysis is needed to understand the actual situation.

\section{Comparison with Other data: One-Dimensional Distributions} \label{sec:comp}

\subsection{$^{12}CO$ (1-0) versus $^{12}CO$ (3-2)} \label{sec:comp_co}

The FUGIN survey mapped part of the first quadrant of the Galactic plane 
at the lowest rotational transition (1-0) of three CO isotopologues ($\mathrm{^{12}CO/^{13}CO/C^{18}O}$) with an angular resolution
comparable to that of the COHRS survey (see also Table~\ref{tab:surveys} for detailed survey information).
\COontoze\ is a fundamental rotation transition expected to be excited 
even in the coldest and most diffuse molecular ISM.
Its critical density at 10\,K is $\sim 10^3~\mathrm{cm^{-3}}$
while that of \COthtotw\ is $\sim 10^4~\mathrm{cm^{-3}}$.
However, for $\mathrm{^{12}CO}$, radiative trapping causes the molecule to emit at a density an order of magnitude lower than its critical density.
In any case, the (3–2) emission line is emitted in relatively warmer and denser ISM conditions than (1–0).

\begin{figure*}
\centering
\includegraphics[width=160mm]{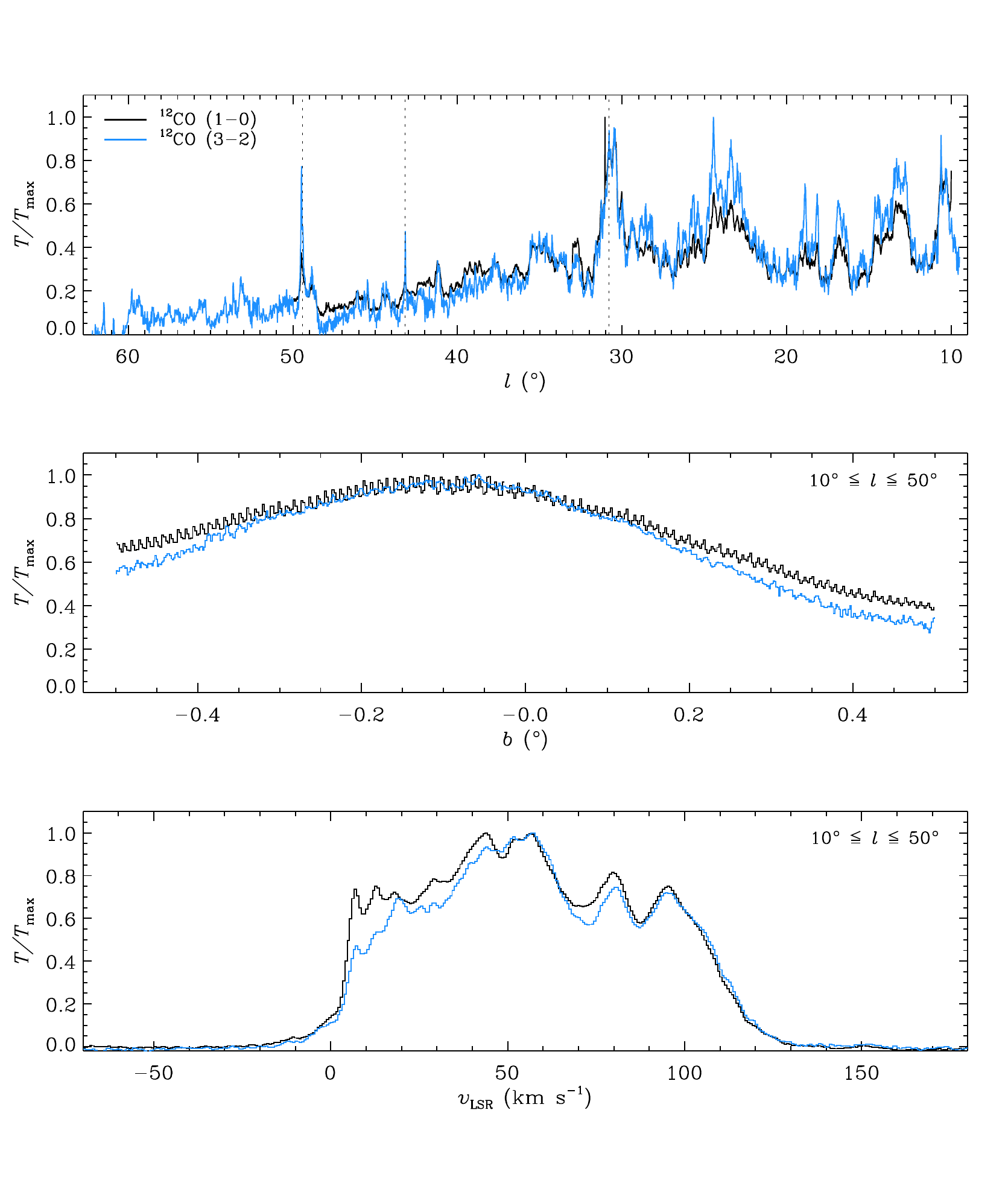}
\caption{
Integrated (one-dimensional) $l$-, $b$-, and $\vlsr$-distributions
of \COontoze\ (FUGIN; black) and \COthtotw\ (COHRS; blue).
Each profile is obtained by integrating across the two orthogonal axes.
Note that $b$- and $v$-profiles are used data within $10\arcdeg \leq l \leq 50\arcdeg$ 
since available FUGIN data are limited to the Galactic longitudinal range.
CO intensity is normalized to the peak value of each profile. 
The two CO $l$-profiles are smoothed to have a bin size of $\sim 60$\arcsec\
while their $b$- and $v$-profiles have a bin size corresponding to 
a latitudinal pixel size and a velocity channel width of each CO data respectively.
On the top panel, three vertical dotted lines are drawn to help locate three star-forming regions: W43 ($l= 30\fdg8$), W49A ($43\fdg2$), and W51 ($49\fdg4$). 
}
\label{fig:1dprofile_CO}
\end{figure*}

Figure~\ref{fig:1dprofile_CO} presents longitudinal ($l$), latitudinal ($b$), and velocity ($\vlsr$) distributions of 
normalized integrated intensity for COHRS \COthtotw\ (blue profiles) and FUGIN \COontoze\ (black profiles).
Each profile was obtained by integrating over the two orthogonal axes,
and then the intensity was normalized to the peak intensity in the profile.
Note that FUGIN was mapped over a more-limited longitudinal range ($l \leq 50\arcdeg$) than COHRS.
Since intensity variations along longitudes are much larger than the pixel size of the survey data
(6\arcsec\ for COHRS and 8\farcs5 for FUGIN),
the original $l$-profiles are smoothed to have a bin size of 60\arcsec\ 
using the IDL INTERPOL function to provide a better visual exhibition.
On the other hand, the $b$- and $\vlsr$-profiles are displayed without smoothing.
Periodic oscillations in the FUGIN $b$-profile exhibit a horizontal stripe pattern caused by instrumental artifacts.

Both $l$-profiles integrated over latitude and velocity for the two transitions
tend to decrease in general above $l = 30\arcdeg$.
It is because the distribution of molecular ISM is not uniform in the Galactic disk
and the Sun is far from the Galactic center. 
First, most of the molecular ISM is concentrated in the inner Galactic disk, 
and the distance ($\Delta d_\mathrm{los}$) from the Sun and the far side of the inner disk steeply decreases with increasing longitude
($\Delta d_\mathrm{los} \sim 6$~kpc for two lines of sight, $l=60\arcdeg$ and 30\arcdeg, 
while $\Delta d_\mathrm{los} \sim 2$~kpc for two lines of sight, $l=30\arcdeg$ and 10\arcdeg). 
In other words, more CO emission lines usually accumulate along the line of sight towards lower Galactic longitudes.
Second, the molecular ISM is strongly associated with Galactic spiral arms 
and the arrangement of the spiral arms along with a line of sight 
affects the shape of the $l$-profiles (see Figure~\ref{fig:lvitg} for an example).
Above $l = 30\arcdeg$, there are fewer spiral arms lying close to the Sun.
On the other hand, $l-$profiles with $l < 30\arcdeg$ show large fluctuations
rather than a smooth change in profile strength.
This is mainly caused by how many luminous velocity components (or GMCs) overlap in a line of sight.
In comparison between the two transition profiles,
the peak locations are generally equal to each other.
The strengths of the two profiles are also similar in some longitudes 
(for example, at $l \sim 30\fdg5$ and 35\arcdeg),
but their distinction is clearly seen in many other longitudes.
While (3-2) emission becomes more strongly peaked than (1-0) emission,
for example, at $l \sim 13\arcdeg$, 23\fdg5, 24\fdg5, 43\arcdeg, and 49\fdg5,
while the predominance of (1-0) emission appears mainly at weak peaks or between peaks.
The three prominent star-forming regions mentioned in Section~\ref{sec:example}
are also well located at CO peaks.
The comparison with such star-forming population will be discussed in the next section.

The $b$-profiles integrated across longitude and velocity for the two CO transitions 
have a shape close to a normal distribution
since lots of CO-emission components are integrated over a wide ($l$, $\vlsr$) range.
Least-square Gaussian fitting gives best fit functions of 
$0.940\,\mathrm{exp}(-\frac{1}{2}(b-0.115)^2/0.428^2)$ and
$0.948\,\mathrm{exp}(-\frac{1}{2}(b-0.105)^2/0.365^2)$ 
for (1-0) and (3-2), respectively.
They have a nearly equal peak position and a slightly-stronger negative wing than a positive one, 
while the normalized intensity of the (3-2) profile decreases more rapidly than that of the (1-0) profile.

As shown in the $l$- and $b$-profiles,
the $\vlsr$-profiles integrated over longitude and latitude for the two transitions
look similar each other.
In other words, the locations of peaks normalized by maximum intensity of each profile are almost the same. 
However, except for the two peaks at $\vlsr \sim 50$--60~$\kms$,
(1-0) emission is always stronger than (3-2) emission.
The difference is relatively more pronounced at $\vlsr \sim 5$--15~$\kms$, indicating the presence of more diffuse local emission in the (1-0).

\subsection{$^{12}CO$ Emission versus Star-forming Population} \label{sec:comp_co_sf}

The detection of \schii\ regions is the clearest evidence for ongoing massive star formation.
The \wise\ catalog provides the entire sky Galactic \schii\ regions identified using mid-infrared data.
In the area where COHRS and FUGIN overlap ($l= 10\arcdeg$--50\arcdeg\ and $b\leq 0\fdg5$),
we found 2179 \wise\ \schii\ regions were found, except for one source without radio data. 
\sout{These are all \wise\ objects in the area except for one source without radio data.}
On the other hand, as the densest parts within GMCs are where star formation can take place,
dense molecular clumps can be at various early evolutionary stages of star formation,
from starless to early embedded stages.
The ATLASGAL compact-source catalog provides about 10,000 dense clumps in the range of $|l| < 60\arcdeg$ and $|b| < 1\fdg5$ \citep{contreras2013, urquhart2014a} found by using submillimeter survey data.
For about 8000 dense clumps of them in $5\arcdeg < |l| < 60\arcdeg$,
\citet{urquhart2018} investigated their detailed properties including velocities and distances, luminosities, and masses and inferred evolutionary stages using mid- and far-infrared survey data.
Their classification scheme divides clumps into four groups:
massive star-forming (MSF) clumps, YSO-forming clumps, protostellar clumps, and starless or pre-stellar clumps.
In the area where COHRS and FUGIN overlap,
2178 ATLASGAL clumps with signs of star-formation 
(i.e., except for those classified as a quiescent phase or unclassified) were identified:
including 455 MSF clumps, 1222 YSO-forming clumps, and 501 protostellar clumps.

\begin{figure*}
\centering
\includegraphics[width=160mm]{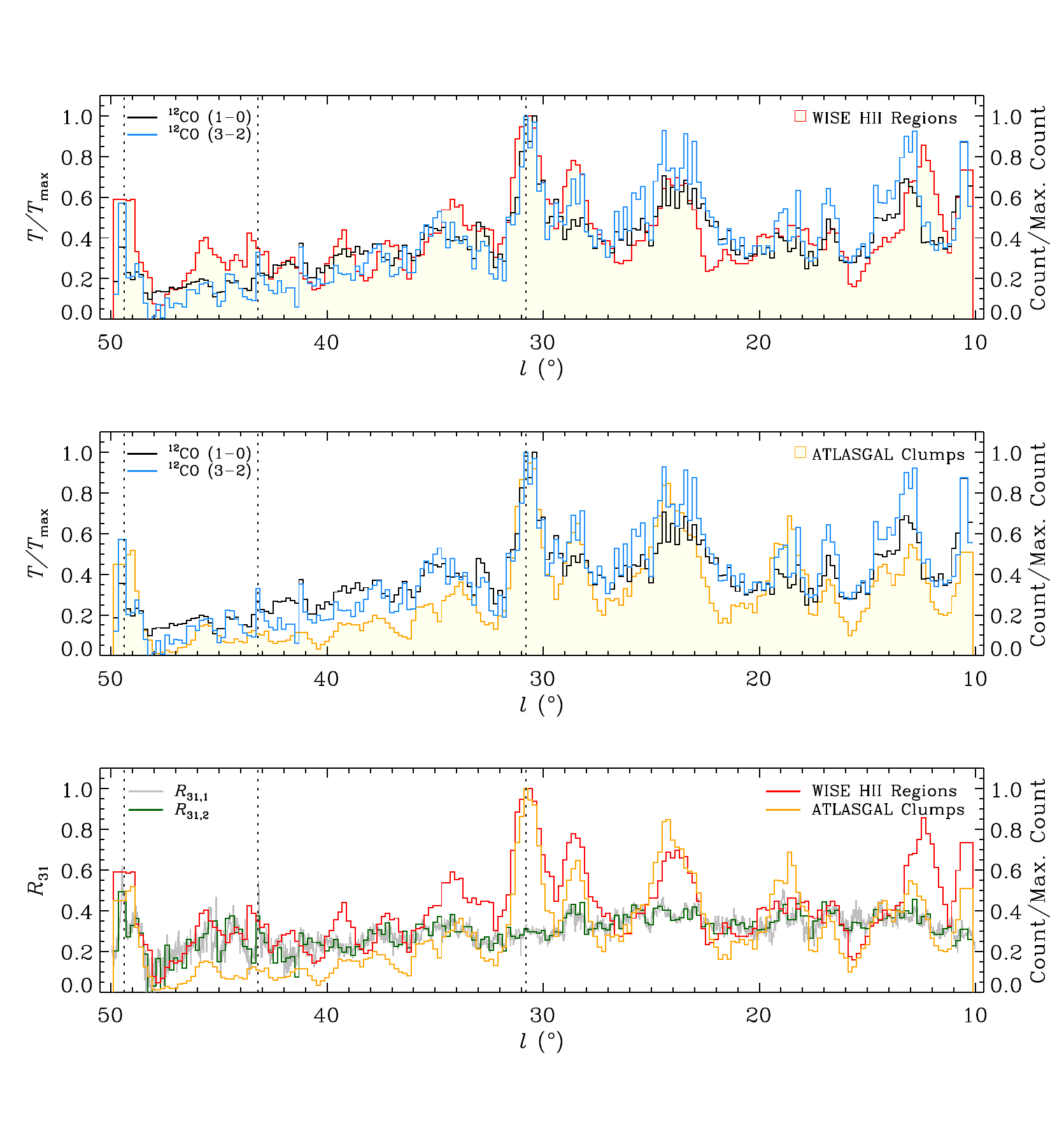}
\caption{
Integrated $l$-distribution
of \COontoze\ (FUGIN; black), \COthtotw\ (COHRS; blue),
Galactic \schii\ regions (\wise; red in the top and bottom panel), 
star-forming clumps (ATLASGAL; orange in the middle and bottom panel),
and the line ratio \Rco\ (gray and green in the bottom panel).
Each profile is integrated over $b\leq 0\fdg5$ and $-60~\kms < \vlsr < +170~\kms$.
The star-forming clumps contain all except starless clumps among the ATLASGAL compact-source catalog, that is, massive star-forming clumps, young stellar object-forming clumps, and protostellar clumps.
CO histograms are drawn in the same way as in Figure~\ref{fig:1dprofile_CO},
but interpolated to have the same bin size (12\arcmin) and abscissa values as those of \schii\ regions or star-forming clumps.
For \schii\ regions and star-forming clumps,
each histogram is obtained by counting the number of sources in a moving bin
with a population-counting bin width of 1\arcdeg\ and a bin-moving interval of 12\arcmin.
The green profile (\Rcotwo) is the ratio of the two CO-transition profiles in the upper panels, but on an absolute scale that is not normalized by maximum. 
Likewise, the gray profile ($R_{31,1}$) is the ratio on the absolute scale of the two CO profiles shown in Figure~\ref{fig:1dprofile_CO}.  
Three vertical dotted lines are drawn to help locate three star-forming regions: W43 ($l= 30\fdg8$), W49A ($43\fdg2$), and W51 ($49\fdg4$).
}
\label{fig:1dprofile_COandSFR}
\end{figure*}

Figure~\ref{fig:1dprofile_COandSFR} shows almost the same normalized $l$-profiles
as shown in Figure~\ref{fig:1dprofile_CO} for the star-forming populations of \wise\ \schii\ regions and ATLASGAL clumps together with FUGIN \COontoze\ and COHRS \COthtotw.
The histograms of the star-forming populations are obtained by counting the number of each catalog source
in the moving bin with a bin size of 1\arcdeg\ and a step size of 12\arcmin.
These moving-bin histograms avoid bias due to a specific bin size.
For statistical comparative analysis, the two CO-emission profiles are interpolated are
using the IDL INTERPOL function
to have a bin size (12\arcmin) equal to that of the star-forming population profiles. 
In the given longitude range, 
all profiles reach their largest peak at the longitudes in the W43 direction ($30\fdg8$).
At the longitudes in the W51 direction ($49\fdg4$), all of them also show a distinct peak.
However, near the W49A ($43\fdg2$) direction, the peak height decreases significantly with smoothing.
At $l=49\fdg5$, \COthtotw, \wise\ \schii\ regions, and ATLASGAL clumps show a huge excess compared to \COontoze, 
which can indicate a very high temperature of CO due to active star-forming activities,
while at $l=30\arcdeg$, all four distributions exhibit similar intensities.

In addition, a broad hump centered around $l \sim 24\arcdeg$ stands out. 
The distribution of \COthtotw\ shows three thin and sharp peaks, one of which coincides with the peak of the ATLASGAL clumps.
However, the distributions of \COontoze\ and \wise\ \schii\ regions are relatively smooth.
The G24\arcdeg-hump is seen as a combination of \wise\ \schii\ regions/GMCs close to each other in the longitude direction \sout{as well as those in the same line of sight}:
for example, GMCs with massive star-forming activities such as G23.01$-$0.41 at $\sim 77~\kms$,  G23.44$-$0.18 at $\sim 100~\kms$, and G25.38$-$0.18 (W42) at $\sim 65~\kms$ \citep[e.g.,][]{brunthaler2009, ohishi2012, su2015, dewangan2015}.
This incredibly rich line-of-sight is being targeted for the GASTON Galactic plane survey \citep{rigby2021}.
Interestingly, the distribution of \wise\ \schii\ regions shows a strong peak at about 12\fdg5, but not the rest.
The peak does not appear when counting only objects, which have a single measured velocity, marked in Figure~\ref{fig:lvitg}.
About half of the \wise\ \schii\ regions contributing to the peak do not have measured velocity information and most of them are classified as radio-quiet sources.
This area might contain many old \wise\ \schii\ regions that have dispersed the molecular clouds they were formed in.

\begin{figure*}
\centering
\includegraphics[width=160mm]{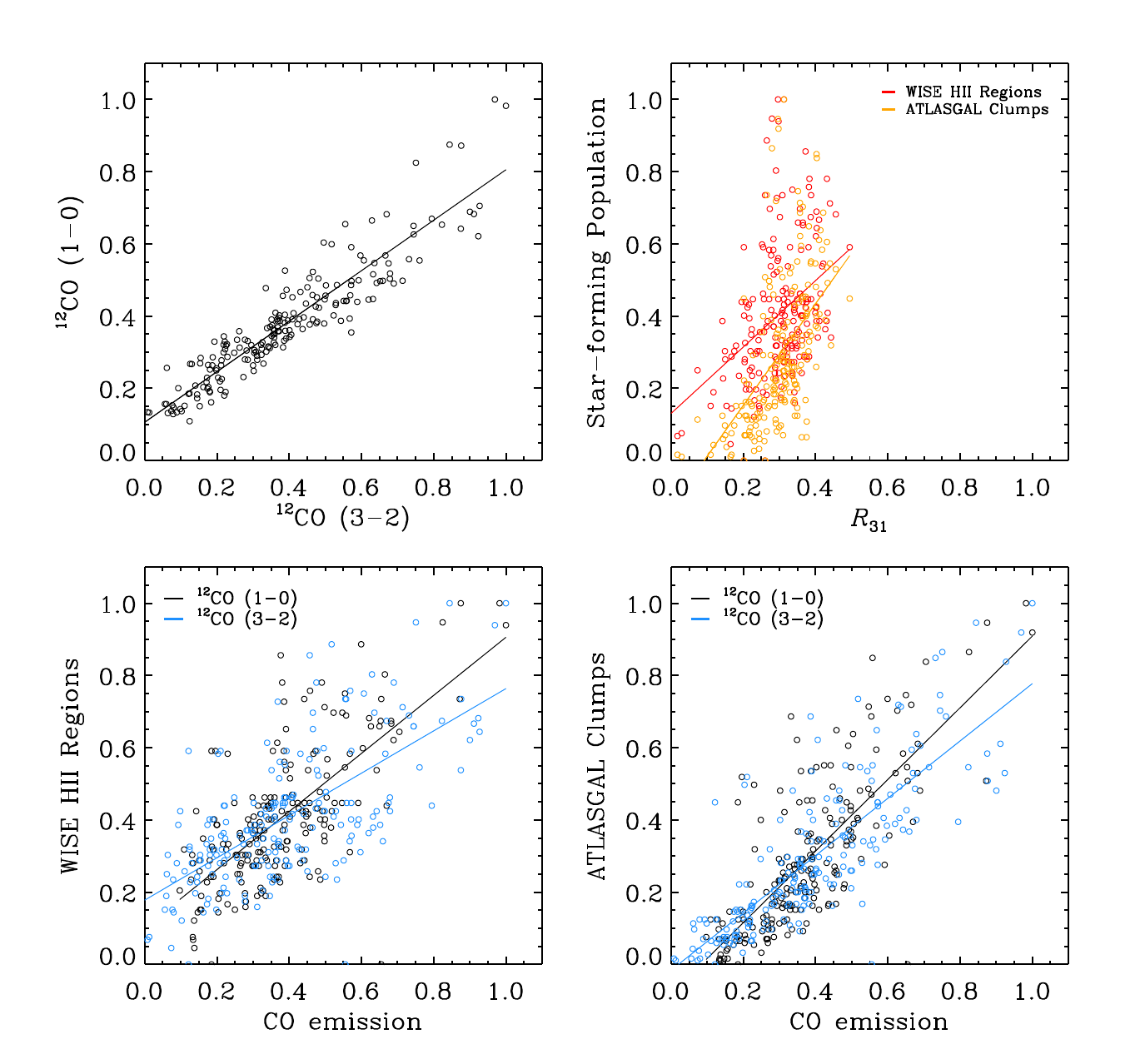}
\caption{
Scatter plots between two targets using the normalized histogram values shown in Figure~\ref{fig:1dprofile_COandSFR}:
clockwise from top left, \COontoze\ vs. \COthtotw, 
\wise\ \schii\ regions (red) or ATLASGAL star-forming clumps (orange) vs. \Rco,
\wise\ \schii\ regions vs. CO emission,
and  star-forming clumps vs. CO emission.
In the two bottom panels, \COontoze\ and \COthtotw\ are displayed with
black and blue circles, respectively.
A solid line is the best result of least-squares fit in a linear model using the IDL LINFIT procedure.
}
\label{fig:1dprofile_COandSFR_corr}
\end{figure*}

We estimated the line ratio of the two CO transitions, i.e., \Rco\,$\equiv$ \COthtotw/\COontoze.
\Rco\ $l$-distribution with two different bin sizes is displayed in the bottom panel of Figure~\ref{fig:1dprofile_COandSFR} and compared to the \wise\ \schii\ regions or ATLASGAL clumps.
For \Rco, a gray profile ($R_{31,1}$) has the same 60\arcsec\ bin size as the CO $l$-profile in Figure~\ref{fig:1dprofile_CO},
and a green profile (\Rcotwo)  has the same 12\arcmin\ bin size as the other profiles in Figure~\ref{fig:1dprofile_COandSFR}.
We find a median \Rco\,$=$ 0.27, which is similar to the mean value of 0.31 for nearby galaxies \citep{leroy2022}.
Compared to the CO profiles normalized by maximum shown in the upper panels, the \Rco\ profile shows less dramatic variation except near the W51 direction (See Figure~\ref{fig:1dprofile_COandSFR_corr} also).
At $l=48\fdg2$, a deep valley is visible, with the higher-longitude side increasing steeper than the lower one.
At $l=49\fdg4$, W51's line of sight, the \Rco\ profile is peaked (\Rcotwo\,$=$ 0.50).

A scatterplot was drawn for each pair, as shown in Figure~\ref{fig:1dprofile_COandSFR_corr}, 
and the Spearman correlation test was applied to evaluate the relationship of results.
We used \texttt{pymccorrelation}\footnote{\url{https://github.com/privong/pymccorrelation}} of \citet{privon2020}, which is a Python implementation and expansion of a Monte Carlo error-analysis procedure described by \citet{curran2014}. 
We computed the Spearman correlation coefficient ($\rho$) using 1000 bootstrapping iterations,
and the median and 16/84 percentile ranges are listed in Table~\ref{tab:corr}.
$\rho = +1$ or $-1$ is a perfect positive or negative correlation while $\rho = 0$ is no correlation between the data.
For the two CO transitions, $\rho$ is 0.94.
It is not surprising that they show a very strong positive correlation.
There is also a positive correlation between CO and star-forming population.
As the ATLASGAL clumps used in this paper were selected only for those that form stars, it is natural that there is a strong positive correlation between \wise\ and ATLASGAL (\wise-ATLASGAL).
On the other hand, ATLASGAL clumps shows a stronger correlation with CO or \Rcotwo\ than \wise\ \schii\ regions. 
That is because that the ATLASGAL catalog contains sources in earlier evolution stages than the \wise\ catalog,
which are still deeply embedded in their natal molecular cloud.
Comparing the relationship between the two transitions and ATLASGAL clumps, the COHRS-ATLASGAL correlation coefficient is greater than the FUGIN-ATLASGAL correlation coefficient.
Also, the difference in correlation coefficient between COHRS-\wise\ and COHRS-ATLASGAL is larger than that between FUGIN-\wise\ and FUGIN-ATLASGAL.
These are explained by \COthtotw\ being more sensitive to dense gas than \COontoze.
In addition, the COHRS-ATLASGAL correlation is slightly stronger than ATLASGAL-\wise.
Thus, the $J=$ (3-2) transition is a better tracer of star-forming gas.

\begin{deluxetable}{l|cccc}
\tabletypesize{\scriptsize}
\tablecaption{Spearman Correlation Coefficients \label{tab:corr}}
\tablewidth{0pt}
\tablehead{
\colhead{} & \colhead{COHRS} & \colhead{FUGIN} & \colhead{\Rcotwo}  & \colhead{ATLASGAL}
}
\startdata
FUGIN            & 0.94\Vectorstack{+0.01 -0.01} & \nodata              & \nodata   &  \nodata \\
\wise            & 0.65\Vectorstack{+0.05 -0.05} & 0.65\Vectorstack{+0.05 -0.05} & 0.42\Vectorstack{+0.06 -0.06} & 0.81\Vectorstack{+0.03 -0.03} \\
ATLASGAL         & 0.84\Vectorstack{+0.03 -0.03} & 0.76\Vectorstack{+0.04 -0.05} & 0.66\Vectorstack{+0.05 -0.05} & \nodata \\
\enddata                   
\tablecomments{50\% value and 16\%/84\% range of coefficient probability distribution obtained by bootstrapping with 1000 iterations.}
\end{deluxetable}



\section{Summary} \label{sec:sum}

We present the full data of COHRS, which is a survey mapping a region of the Galactic plane, covering $9\fdg5 \le l \le 62\fdg3$ and $|b| \le 0\fdg5$, in \COthtotw\ using HARP on the JCMT.
Since the initial public release of \dempseypaper, further observations have been made to reach the full scope of the survey, and improved data-reduction processes have been applied, including steps to mitigate off-position contamination effects.
The COHRS data are publicly accessible at \dataset[doi:10.11570/22.0078]{\doi{10.11.570/22.0078}}.
The released data have an angular resolution of 16\farcs6 and a velocity resolution of 0.635~$\kms$ with a velocity coverage of $-200~\kms < \vlsr < +300~\kms$.
The data are sampled on 6\arcsec\ pixels and achieve a mean RMS of 0.6\,K.

We investigate integrated one-dimensional distribution of COHRS \COthtotw\ and compare with those of FUGIN \COontoze\ or star-forming population (\wise\ \schii\ regions and ATLASGAL star-forming clumps).
When comparing them in the integrated longitudinal space, the peak locations are generally similar to each other, but differences in peak intensity can be seen in many longitudes.
For example, the distinct peak of $l = 12\fdg5$, visible only in \wise\ \schii\ regions, suggests that old star-formation regions are distributed in the line of sight and the surrounding molecular gas has already blown away.
All available pairs (\COontoze\ vs. \COthtotw\ and
\wise\ \schii\ regions or ATLASGAL star-forming clumps vs. \Rco\ or CO emission)
represent a positive correlation.
The relationship between \COthtotw\ and ATLASGAL clumps is slightly stronger than that between \COthtotw\ and \wise\ \schii\ regions, while the relationship between \COontoze\ and the two star-formation tracers is relatively similar.
This can happen because the higher CO transition traces denser areas within molecular clouds and are more closely related to early star-formation stages.

The COHRS data will be complemented with existing and upcoming CO and continuum surveys to study statistical properties of molecular gases along the Galactic plane as well as detailed structures and properties of individual objects. 
These high-resolution data of molecular gas will also help to investigate outflow activities in star-forming regions.
Methanol masers are an unambiguous indicator of massive star formation, and \citet{green2010} and \citet{breen2015} provide unbiased 6\,GHz class II methanol maser surveys in the COHRS area.
In future work, we will investigate outflow features toward massive star-formation regions where methanol masers are detected.


\clearpage
\acknowledgments

G.P. is supported by Basic Science Research Program through the National Research Foundation of Korea (NRF) funded by the Ministry of Education (NRF-2020R1A6A3A01100208).
This work was partly supported by the Korea Astronomy and Space Science Institute grant funded by the Korea government(MSIT) (Project No. 2022-1-840-05).

The James Clerk Maxwell Telescope is operated by the East Asian Observatory on behalf of The National Astronomical Observatory of Japan; Academia Sinica Institute of Astronomy and Astrophysics; the Korea Astronomy; the National Astronomical Research Institute of Thailand and Space Science Institute; Center for Astronomical Mega-Science (as well as the National Key R\&D Program of China with No. 2017YFA0402700). Additional funding support is provided by the Science and Technology Facilities Council of the United Kingdom and participating universities and organizations in the United Kingdom and Canada.

The James Clerk Maxwell Telescope has historically been operated by the Joint Astronomy Centre on behalf of the Science and Technology Facilities Council of the United Kingdom, the National Research Council of Canada and the Netherlands Organisation for Scientific Research.

The authors wish to recognize and acknowledge the very significant cultural role and reverence that the summit of Maunakea has always had within the indigenous Hawaiian community.  We are most fortunate to have the opportunity to conduct observations from this mountain.

MJC thanks Peter Chiu of RAL Space for obtaining the hardware used for the COHRS
data reduction, and for ongoing computer-troubleshooting support.

This publication makes use of data from FUGIN, FOREST Unbiased Galactic plane Imaging survey with the Nobeyama 45-m telescope, a legacy project in the Nobeyama 45-m radio telescope.

\facilities{JCMT}
\software{The IDL Astronomy User’s Library \citep{landsman1993},
          numpy \citep{walt2011},
          pymccorrelation \citep{curran2014, privon2020},
          scipy \citep{virtanen2020},
          Starlink \citep{currie2014},
          SWarp \citep{bertin2002}.
          }


\appendix
\restartappendixnumbering

\section{Parameters for Quality Assurance} \label{app:qa}

For an observation to be included in the group phase of the
\textsc{ORAC-DR} reduction, a number of quality-assurance criteria have
to be met.  These are listed below. Brief descriptions of the parameters may be found in \citet[Appendix H]{Thomas2018}.

\begin{verbatim}
BADPIX_MAP = 0.3
GOODRECEP = 10
TSYSBAD = 2000
FLAGTSYSBAD = 0.5
TSYSMAX = 1500
TSYSVAR = 1.0
RMSVAR_RCP = 1.0
RMSVAR_SPEC = 0.4
RMSVAR_MAP = 2.0
RMSTSYSTOL = 0.5
RMSTSYSTOL_QUEST = 0.15
RMSTSYSTOL_FAIL = 0.2
RMSMEANTSYSTOL = 1.0
CALPEAKTOL = 0.2
CALINTTOL = 0.2
RESTOL = 1
RESTOL_SM = 1
\end{verbatim}

Given the COHRS data were observed in poor conditions,
some with 225\,GHz opacity greater than 0.3, we chose a relaxed
maximum $T_\mathrm{sys}$ to be more inclusive, to permit noisy
data to be combined with repeat observations of similar quality.

\section{Example Recipe-Parameter File} \label{app:recpar}

Fine control of \textsc{ORAC-DR} recipes may be achieved through
recipe parameters. Every region observed in COHRS has an associated
recipe file.  Below is an annotated example, re-ordered for
convenience. Boolean parameters are assigned 1 for true and 0 for
false. Spectral-channel ranges are all measured in $\kms$.

\begin{verbatim}
[REDUCE_SCIENCE_NARROWLINE]
#
# MAKECUBE parameters
#
CUBE_WCS = GALACTIC
PIXEL_SCALE = 6.0
SPREAD_METHOD = gauss
SPREAD_WIDTH = 9
SPREAD_FWHM_OR_ZERO = 6
#
# Tiling and chunking
#
TILE = 0
CHUNKSIZE = 12288
CUBE_MAXSIZE = 1536
#
# Baseline
#
BASELINE_ORDER = 1
\end{verbatim}

The above apply to the \textsc{REDUCE\_\-SCIENCE\_\-NARROWLINE} recipe,
and are constant for all the parameter files. The first stanza
controls how the PPV spectral cubes are sampled as described in
Section~\ref{sec:dr}. The middle group was not essential, but it
allowed the PPV cube to be made as one, rather than fragmenting into
chunks. The final parameter defined the baseline-fitting polynomial
order. See Section~\ref{sec:dr}.

\begin{verbatim}
#
# Bad-baseline filtering
#
BASELINE_LINEARITY = 1
BASELINE_LINEARITY_LINEWIDTH = -25:87
BASELINE_LINEARITY_MINRMS = 0.080
HIGHFREQ_INTERFERENCE = 1
HIGHFREQ_RINGING = 0
HIGHFREQ_INTERFERENCE_THRESH_CLIP = 4.0
LOWFREQ_INTERFERENCE = 1
\end{verbatim}

These parameters decided which tests were performed on the raw data to
reject spectra containing significant non-astronomical signal. All
observations' processing tested that baselines were not grossly deviant
from linearity, both for individual spectra
(\texttt{LOWFREQ\_INTERFERENCE}) and for each receptor as a whole
(\texttt{BASELINE\_LINEARITY}).  The QA also hunted for spectra with
alternating abnormally bright and dark fluxes
(\texttt{HIGHFREQ\_INTERFERENCE}).  \texttt{HIGHFREQ\_RINGING} was
only enabled if ringing \citep{jenness2015} was detected in Receptor
H07. \texttt{BASELINE\_LINEARITY\_LINEWIDTH} specified a velocity range
to exclude from the non-linearity tests.  When the astronomical
emission is not restricted to a single range, we set
\texttt{BASELINE\_LINEARITY\_LINEWIDTH = base} to request that
\texttt{BASELINE\_REGIONS} be used instead to define the
velocity ranges to \emph{include} in the tests.
\texttt{BASELINE\_LINEARITY\_MINRMS} was the minimum RMS deviation
from linearity, measured in antenna temperature, for a receptor to be
flagged as bad.  Well-behaved data had RMS values range from 0.01 to
0.03. \texttt{HIGHFREQ\_INTERFERENCE\_THRESH\_CLIP} set the number of
standard deviations at which to threshold the noise profile of raw
spectra above its median level, in order to decide whether to reject
spectra with high-frequency noise.

\begin{verbatim}
#
# Flatfield receptors
#
FLATFIELD = 1
FLAT_METHOD = sum
FLAT_REGIONS = 12.0:25.2,27.2:29.1,35.7:41.5
\end{verbatim}

These parameters defined whether or not to flat-field
(\texttt{FLATFIELD}); always with the summation method
\texttt{FLAT\_METHOD}, which proved to be the most stable; and the list
of velocity ranges over which to integrate the fluxes for each
receptor. If \texttt{FLATFIELD = 0}, the subsequent flat-field
parameters were ignored.

\begin{verbatim}
#
# Reference-spectrum removal from time-series cubes
#   - Automatic
#
SUBTRACT_REF_EMISSION = 1
CLUMP_METHOD = clumpfind
REF_EMISSION_MASK_SOURCE = both
REF_EMISSION_COMBINE_REFPOS = 1
REF_EMISSION_BOXSIZE = 19
#
#   - Manual location
#
SUBTRACT_REF_SPECTRUM = 1
REF_SPECTRUM_COMBINE_REFPOS = 1
REF_SPECTRUM_REGIONS =-1.5:0.1,2.5:4.0,7.0:11.1
\end{verbatim}

The reference (off) position for the majority of the observed regions
contained emission that appears as absorption features in all spectra.
When detected after inspection of the first-pass reductions PPV cubes,
the removal techniques were enabled by switching on
\texttt{SUBTRACT\_REF\_SPECTRUM}.  We did not want any unnecessary
modification of spectra where no evident off-position was visible.

An outline of the methods used can be found in
Section~\ref{sec:off-pos}. The first stanza defined parameters for the
automated method. \texttt{REF\_EMISSION\_MASK\_SOURCE} used not only
used the source-masked spectrum to locate the lines, but also the
unmasked modal spectrum to determine the line strengths.  The emission
was located with the ClumpFind algorithm \citep{williams2011}
applied in one dimension by \textsc{FINDCLUMPS} from the
\textsc{CUPID} \citep{berry2007} package. In rare circumstances
where repeat observations had switched reference positions, each
reference position was analysed separately. (\texttt{REF\_EMISSION\_COMBINE\_REFPOS}).

The second stanza was to deal with residual off-position signal, that
the automated method left, being either untouched lines or, most
commonly, lines reduced in depth but not eliminated. Application of
this algorithm was enabled by \texttt{SUBTRACT\_REF\_SPECTRUM}.
\texttt{REF\_SPECTRUM\_COMBINE\_REFPOS} performed the equivalent
action as \texttt{REF\_EMISSION\_COMBINE\_REFPOS}. The list of the line
extents were supplied through \texttt{REF\_SPECTRUM\_REGIONS}.

In the seven cases where even the manual guidance did not remove all
the absorption lines, the name of a manually determined off-position
residual spectrum was supplied through \texttt{REF\_SPECTRUM\_FILE}
(not shown above).

\begin{verbatim}
#
# Properties of final products
#
FINAL_LOWER_VELOCITY = -230
FINAL_UPPER_VELOCITY = 355
REBIN = 0.635,1.0
\end{verbatim}

The velocity limits of the PPV cubes were set by the first two recipe
parameters. These limits were further trimmed during mosaic formation
in order to prevent exceeding the maximum number of array elements.
\texttt{REBIN} assigned velocity resolutions for re-gridded PPV cubes,
generating one at the R2 width of 0.635\,$\kms$, and the other at
1.0\,$\kms$ width for comparison with R1.

\begin{verbatim}
#
# Moment maps
#
MOMENTS_LOWER_VELOCITY = -43
MOMENTS_UPPER_VELOCITY = 84
LV_IMAGE = 1
LV_AXIS = skylat
LV_ESTIMATOR = sum
\end{verbatim}

For completeness, the final set of parameters asked for the creation
of a longitude-velocity (LV) map, summing over galactic latitude.
These LV maps were for a quick inspections of the reductions, and do not
form part of the release. The released LV maps were derived from the
mosaics. The first two parameters restrict the velocity range when
computing the moments map, and were used for efficiency.

\section{Automated algorithm for removal of Off-position Signal} \label{app:off-pos}

This appendix expands on the outline, presented in Section~\ref{sec:off-pos},
of the automated algorithm to remove off-position signals.

\begin{enumerate}

\item Data observed at different reference positions are processed
separately.

\item The initial step is to collapse the time axis by forming the
mode at each spectral channel. Those modal spectra are mildly smoothed
with a 1.5-channel full-width half-maximum Gaussian point-spread
function in order to define the extents of the absorption lines
better.  The mode at each spectral channel was determined by an
iterative maximum-likelihood function, for which the data were 
inversely weighted by their deviations from the current mean.
At each iteration outliers at 3.0 standard deviations from the current
mean were clipped.  Iterations proceeded until convergence to a
stationary point.

\item Refinement of the modal spectra occurs for each receptor as follows.

\begin{enumerate}
\item The lines under analysis are always in emission, as required by
the clump-finding software.

\item Before the locations of reference-spectrum emission lines are
determined within the masked-source modal spectrum, an attempt to
remove residual source emission is made. Its steps are: subtract a
75-pixel median-smoothed version, then mask channels that fall below a
$-3 * \mathrm{rms}$ threshold, then repeat the first step but having
the kernel reduced to 41 pixels.

\item In the search for off-position lines the background is not
initially subtracted.  While this choice may lose weaker reference
emission embedded in extended source signal, it compensates by not
regarding dips in the source signal as reference emission.

\item Line properties come from \textsc{CUPID}'s \textsc{FINDCLUMPS} with 
a tuned ClumpFind method, with $2 * \mathrm{rms}$ minimum detection level.
Consequently, to allow for the wings near the baseline, an additional
three pixels either side of the line masked.  A fixed 19-channel
smoothing kernel is used to determine the background for the line finding
(but there is an option to measure the widest line iteratively in order
to set the smoothing kernel).

\item The masked channels for the reference and the source are applied to
each unmasked modal spectrum, which is analysed in the same fashion 
as for the masked modal spectrum.

\item Any residual background from spectrally broad source emission is
removed with \textsc{FINDBACK} from the \textsc{CUPID} package once
the masked channels are filled using an iterated solution to
Laplace's Equation.  The revised background is more accurate as the
bulk of the emission and off-position lines have been excised.

\item Measure the properties of reference lines once again, now
improved by the more-accurate background.

\item Remove any varying residual background to cater for spectrally
extended source emission.  Use a narrow (9-channel) kernel to track the
background more precisely.

\item A bias remains in the background subtraction and a $1.5 *
\mathrm{rms}$ empirical correction is added.

\item Masked values beyond the spectral lines in the estimated reference
spectrum are set to zero.

\end{enumerate}

\item For data taken at different epochs, the mapping from pixel to
velocity is likely to be different, so they are aligned to the first
epoch.  This permits pixel-by-pixel subtraction.

\item The estimated reference spectrum is expanded to the bounds of the
raw time series, from which the spectrum is subtracted.
\end{enumerate}

\section{Average RMS noise levels of COHRS tiles} \label{app:avg_rms}

\startlongtable
\begin{deluxetable}{rlc}
\tabletypesize{\scriptsize}
\setlength{\tabcolsep}{2mm}
\tablewidth{0pt}
\tablecaption{COHRS Tiles Summary \label{tab:tiles}}
\tablehead{
\colhead{\#} & \colhead{Tile Name\tablenotemark{\footnotesize a}} & \colhead{RMS Noise\tablenotemark{\footnotesize b}} \vspace{-2mm}\\
\colhead{} & \colhead{} & \colhead{(K)} 
}
\startdata
  1 & COHRS\_09p50\_0p00 & 0.84 \\
  2 & COHRS\_10p00\_0p00 & 0.72 \\
  3 & COHRS\_10p50\_0p00 & 0.33 \\
  4 & COHRS\_11p00\_0p00 & 0.72 \\
  5 & COHRS\_11p50\_0p00 & 0.33 \\
  6 & COHRS\_12p00\_0p00 & 0.38 \\
  7 & COHRS\_12p50\_0p00 & 0.52 \\
  8 & COHRS\_13p00\_0p00 & 0.37 \\
  9 & COHRS\_13p50\_0p00 & 0.63 \\
 10 & COHRS\_14p00\_0p00 & 0.36 \\
 11 & COHRS\_14p50\_0p00 & 0.51 \\
 12 & COHRS\_15p00\_0p00 & 0.31 \\
 13 & COHRS\_15p50\_0p00 & 0.42 \\
 14 & COHRS\_16p00\_0p00 & 0.39 \\
 15 & COHRS\_16p50\_0p00 & 0.37 \\
 16 & COHRS\_17p00\_0p00 & 0.70 \\
 17 & COHRS\_17p50\_0p00 & 0.90 \\
 18 & COHRS\_18p00\_0p00 & 0.74 \\
 19 & COHRS\_18p50\_0p00 & 0.64 \\
 20 & COHRS\_19p00\_0p00 & 0.85 \\
 21 & COHRS\_19p50\_0p00 & 0.78 \\
 22 & COHRS\_20p00\_0p00 & 0.61 \\
 23 & COHRS\_20p50\_0p00 & 0.43 \\
 24 & COHRS\_21p00\_0p00 & 0.65 \\
 25 & COHRS\_21p50\_0p00 & 0.75 \\
 26 & COHRS\_22p00\_0p00 & 0.76 \\
 27 & COHRS\_22p50\_0p00 & 0.72 \\
 28 & COHRS\_23p00\_0p00 & 0.58 \\
 29 & COHRS\_23p50\_0p00 & 0.53 \\
 30 & COHRS\_24p00\_0p00 & 0.56 \\
 31 & COHRS\_24p50\_0p00 & 0.52 \\
 32 & COHRS\_25p00\_0p00 & 0.52 \\
 33 & COHRS\_25p50\_0p00 & 0.40 \\
 34 & COHRS\_26p00\_0p00 & 0.47 \\
 35 & COHRS\_26p50\_0p00 & 0.61 \\
 36 & COHRS\_27p00\_0p00 & 0.64 \\
 37 & COHRS\_27p50\_0p00 & 0.53 \\
 38 & COHRS\_28p00\_0p00 & 0.46 \\
 39 & COHRS\_28p50\_0p00 & 0.52 \\
 40 & COHRS\_29p00\_0p00 & 0.47 \\
 41 & COHRS\_29p50\_0p00 & 0.46 \\
 42 & COHRS\_30p00\_0p00 & 0.60 \\
 43 & COHRS\_30p50\_0p00 & 0.50 \\
 44 & COHRS\_31p00\_0p00 & 0.68 \\
 45 & COHRS\_31p50\_0p00 & 0.76 \\
 46 & COHRS\_32p00\_0p00 & 0.56 \\
 47 & COHRS\_32p50\_0p00 & 0.50 \\
 48 & COHRS\_33p00\_0p00 & 0.44 \\
 49 & COHRS\_33p50\_0p00 & 0.53 \\
 50 & COHRS\_34p00\_0p00 & 0.46 \\
 51 & COHRS\_34p50\_0p00 & 0.47 \\
 52 & COHRS\_35p00\_0p00 & 0.42 \\
 53 & COHRS\_35p50\_0p00 & 0.43 \\
 54 & COHRS\_36p00\_0p00 & 0.44 \\
 55 & COHRS\_36p50\_0p00 & 0.51 \\
 56 & COHRS\_37p00\_0p00 & 0.50 \\
 57 & COHRS\_37p50\_0p00 & 0.44 \\
 58 & COHRS\_38p00\_0p00 & 0.49 \\
 59 & COHRS\_38p50\_0p00 & 0.44 \\
 60 & COHRS\_39p00\_0p00 & 0.56 \\
 61 & COHRS\_39p50\_0p00 & 0.57 \\
 62 & COHRS\_40p00\_0p00 & 0.49 \\
 63 & COHRS\_40p50\_0p00 & 0.52 \\
 64 & COHRS\_41p00\_0p00 & 0.48 \\
 65 & COHRS\_41p50\_0p00 & 0.58 \\
 66 & COHRS\_42p00\_0p00 & 0.76 \\
 67 & COHRS\_42p50\_0p00 & 0.66 \\
 68 & COHRS\_43p00\_0p00 & 0.57 \\
 69 & COHRS\_43p50\_0p00 & 0.21 \\
 70 & COHRS\_44p00\_0p00 & 0.48 \\
 71 & COHRS\_44p50\_0p00 & 0.50 \\
 72 & COHRS\_45p00\_0p00 & 0.54 \\
 73 & COHRS\_45p50\_0p00 & 0.52 \\
 74 & COHRS\_46p00\_0p00 & 0.59 \\
 75 & COHRS\_46p50\_0p00 & 0.61 \\
 76 & COHRS\_47p00\_0p00 & 0.70 \\
 77 & COHRS\_47p50\_0p00 & 0.82 \\
 78 & COHRS\_48p00\_0p00 & 0.78 \\
 79 & COHRS\_48p50\_0p00 & 0.59 \\
 80 & COHRS\_49p00\_0p00 & 0.71 \\
 81 & COHRS\_49p50\_0p00 & 0.67 \\
 82 & COHRS\_50p00\_0p00 & 0.76 \\
 83 & COHRS\_50p50\_0p00 & 0.62 \\
 84 & COHRS\_51p00\_0p00 & 0.58 \\
 85 & COHRS\_51p50\_0p00 & 0.38 \\
 86 & COHRS\_52p00\_0p00 & 0.73 \\
 87 & COHRS\_52p50\_0p00 & 1.01 \\
 88 & COHRS\_53p00\_0p00 & 0.61 \\
 89 & COHRS\_53p50\_0p00 & 0.80 \\
 90 & COHRS\_54p00\_0p00 & 0.48 \\
 91 & COHRS\_54p50\_0p00 & 0.58 \\
 92 & COHRS\_55p00\_0p00 & 0.56 \\
 93 & COHRS\_55p50\_0p00 & 0.67 \\
 94 & COHRS\_56p00\_0p00 & 0.52 \\
 95 & COHRS\_56p50\_0p00 & 0.47 \\
 96 & COHRS\_57p00\_0p00 & 0.72 \\
 97 & COHRS\_57p50\_0p00 & 0.46 \\
 98 & COHRS\_58p00\_0p00 & 0.67 \\
 99 & COHRS\_58p50\_0p00 & 0.89 \\
100 & COHRS\_59p00\_0p00 & 0.94 \\
101 & COHRS\_59p50\_0p00 & 0.50 \\
102 & COHRS\_60p00\_0p00 & 0.49 \\
103 & COHRS\_60p50\_0p00 & 0.77 \\
104 & COHRS\_61p00\_0p00 & 0.73 \\
105 & COHRS\_61p50\_0p00 & 0.99 \\
106 & COHRS\_62p00\_0p00 & 0.81 \\
\enddata                   
\tablenotetext{a}{The numbers in the tile name give the central longitude and latitude of the tile.}
\tablenotetext{b}{The mean $\TAstar$ RMS noise in the tiles rebinned to 0.635~$\kms$ channel width.}
\end{deluxetable}

\end{document}